\documentclass[fleqn,usenatbib]{mnras}

\usepackage{newtxtext,newtxmath}

\usepackage[T1]{fontenc}
\usepackage{subcaption}
\usepackage{lipsum}

\DeclareRobustCommand{\VAN}[3]{#2}
\let\VANthebibliography\thebibliography
\def\thebibliography{\DeclareRobustCommand{\VAN}[3]{##3}\VANthebibliography}

\usepackage{graphicx}	
\usepackage{amsmath}	
\usepackage[fleqn]{nccmath}

\newcommand{\cm}{\,{\rm cm}}    
\newcommand{\km}{\,{\rm km}}    
\newcommand{\pc}{\,{\rm pc}}     
\newcommand{\kpc}{\,{\rm kpc}}  

\newcommand{\Msun}{\,{\rm M}_{\odot}} 

\newcommand{\s}{\,{\rm s}}      
\newcommand{\Myr}{\,{\rm Myr}} 
\newcommand{\Gyr}{\,{\rm Gyr}}  

\newcommand{\kms}{\km\s^{-1}}    

\newcommand{\muG}{\,\mu{\rm G}} 

\newcommand{\Rvir}{R_{\rm vir}}
\newcommand{\nH}{n_{\rm H}}
\newcommand{\Vrad}{V_{\rm rad}}
\newcommand{\pVrad}{V_{\rm rad}^{v}}
\newcommand{\pV}{V^{v}}
\newcommand{\pVone}{V_1^{v}}
\newcommand{\pVtwo}{V_2^{v}}
\newcommand{\prone}{r_1^{v}}
\newcommand{\prtwo}{r_2^{v}}
\newcommand{\spread}{\Delta_\mathrm{dye}}
\newcommand{\spreadx}{\Delta_\mathrm{dye, \mathit{x}}}
\newcommand{\spready}{\Delta_\mathrm{dye, \mathit{y}}}
\newcommand{\spreadz}{\Delta_\mathrm{dye, \mathit{z}}}

\newcommand{\vecsymmtrmin}{\vec{\mathcal{S}}_{ij, \mathrm{min}}^*}
\newcommand{\vecsymmtrmax}{\vec{\mathcal{S}}_{ij, \mathrm{max}}^*}
\newcommand{\vecsymmtrsum}{\vec{\mathcal{S}}_{ij, \mathrm{sum}}^*}
\newcommand{\vecantisymmreal}{\vec{\Omega}_{ij, \mathrm{real}}}
\newcommand{\vecantisymmsum}{\vec{\Omega}_{ij, \mathrm{sum}}}
\newcommand{\rs}{r_{\rm s}}
\newcommand{\vdis}{\sigma_{\rm vel}}
\newcommand{\KEdens}{{KE}_{\rm dens}}
\newcommand{\tnaught}{t_{\rm 0}}

\newcommand{\rhoover}{\rho_{\rm over}}
\newcommand{\K}{\,{\rm K}}      

\usepackage{hyperref}

\graphicspath{{./}{figures/}}

\newcommand\rev[1]{#1}

\title[Gas mixing in the CGM]{Understanding gas mixing in the circumgalactic medium}

\author[Shah et al.]{
Hilay Shah,$^{1, 2}$\thanks{E-mail: \href{mailto:hilay.shah@anu.edu.au}{hilay.shah@anu.edu.au}}
Freeke van de Voort,$^{2}$
Amit Seta$^{1}$
and Christoph Federrath$^{1, 3}$
\\
$^{1}$Research School of Astronomy and Astrophysics, Australian National University, Canberra, ACT 2611, Australia\\
$^{2}$School of Physics \& Astronomy, Cardiff University, Queens Buildings, The Parade, Cardiff CF24 3AA, UK\\
$^{3}$Australian Research Council Centre of Excellence in All Sky Astrophysics (ASTRO3D), Canberra, ACT 2611, Australia
}

\date{Accepted XXX. Received YYY; in original form ZZZ}

\pubyear{2015}

\begin{document}
\label{firstpage}
\pagerange{\pageref{firstpage}--\pageref{lastpage}}
\maketitle

\begin{abstract}
We study gas mixing in a simulated Milky Way-mass galaxy's circumgalactic medium (CGM) using cosmological `zoom-in' simulations. \rev{We} insert tracer dyes in the CGM \rev{with different gas flows (shearing, coherent, and static) and diverse physical properties to track gas mixing.} We correlate the extent and shape of the dye spread with the local gas properties to understand gas mixing. \rev{Velocity dispersion and traceless symmetric shear tensors (pure shear deformation) in small regions ($\lesssim 5~\rm kpc$) around the dye injection locations best predict the dye spread extent after $200 \Myr$. We use this to determine diffusion calibration constants for subgrid-scale mixing models.} \rev{While the dye shape after} $200 \Myr$ \rev{aligns well with the velocity dispersion and magnetic field dispersion, the best alignment occurs with the dispersion of stretching eigenvectors (traceless symmetric shear tensor) and plane-of-rotation (antisymmetric shear or vorticity tensor) in large regions ($\gtrsim 10 \kpc$) around the dye injection locations. Therefore, shear statistics and velocity dispersion best predict the extent and shape of mixed gas.} The linear temporal dependence of the dye spread suggests superdiffusion in the CGM,
potentially due to turbulent and large-scale coherent flows or numerical diffusion. Despite significant numerical mixing from our 1\,kpc resolution (insufficient to resolve Reynolds numbers $\sim 10^2 \text{–} 10^3$, which require a few hundred pc resolution), our correlation results are robust thanks to fixed spatial resolution throughout the CGM. These results can be used
to predict diffusion coefficients to model magnetic field diffusion, heat transport, and metal mixing.

\end{abstract}




\section{Introduction}
The multiphase circumgalactic medium (CGM) links the interstellar medium (ISM) of galaxies with the diffuse intergalactic medium (IGM). This gas reservoir extends out to about the virial radius of the dark-matter-dominated halo ($\sim 200 \kpc$ for a Milky Way-mass system) and interacts with the enriched outflows from star-forming disks and pristine inflows from the IGM. The cool gas flows accreting from the CGM to the ISM provide the fuel for future star formation. The CGM also contains most of the metals expelled from the ISM. It is thus key to answering questions of sustained star formation \citep{Heavens2004, Bigiel2011, Kennicutt2012, Putman2017}, missing baryons \citep{McGaugh2009, Behroozi2010}, and missing metals in galaxies \citep{Tremonti2004, Peeples2014}. The physical properties of the CGM indirectly influence feedback processes, affecting how galaxies evolve and the resulting outflows \citep{TumlinsonEA2017, CrainandvandeVoort2023, FaucherGigure2023}. Therefore, it is vital to understand its properties and underlying physical processes to understand and test theories of galaxy formation and evolution and interpret observations \citep{TumlinsonEA2017}.

Various physical processes around the CGM make it a multiphase medium. Some major mechanisms include the supernovae-driven hot outflows from the galaxy \citep{Thompson2015, FieldingEA2017}, the cool inflows from the IGM or thermal instabilities \citep{mccourt2012}, mixing between different layers of the cold clouds ejected by galactic wind \citep{Begelman1990, Gronnow2018, Das2023}, cooling induced by stripping of massive satellites \rev{\citep{Roy2024}}\rev{, and cool gas swept up in wakes of satellites \citep{Saeedzadeh2023}}. The presence of diffuse hot coronae (T$\sim 10^6 \K$) has been theoretically predicted by cosmological theories \citep{Spitzer1956, Fukugita1998, Fukugita2006, Stinson2012}. Deep X-ray absorption observations have revealed warm, warm-hot, and hot phases of the CGM \citep{Das2021}. The O\,\textsc{vi}, O\,\textsc{vii}, and O\,\textsc{viii} absorption and emission lines provide evidence for this hot component \citep{Tumlinson2011, Miller2015, Locatelli2024}. The hot phase can also be indirectly probed by the interaction of the Magellanic stream with the hot coronal Milky-Way (MW) gas \citep{Mastropietro2005, Gaensler2008, Fox2010}. The cold component (T$\le 10^4 \K$) is present in the form of high-velocity clouds, detected mainly by H\,\textsc{i}, Si\,\textsc{ii}, Mg\,\textsc{ii} observations \citep{Wakkar1997, Wakkar2007, Shull2009, Steidel2010, Kacprzak2011, Werk2014}. Through cooling of hot gas or mixing of the interface of these clouds with the hot coronal surroundings warm ionised gas is generated (T$\sim 10^5 \K$) \citep{Kwak2011, Armillotta2017}.

\rev{Gas turbulence causes mixing and perturbations of density and temperatures, which influences CGM's multiphase dynamics \citep{Seta&Federrath2022, Das2023}. Understanding turbulent mixing in the CGM can thereby elucidate a variety of gas interactions and properties -- inflows and outflows mixing with the CGM \citep{Pakmor2020, FG2023}, mass outflow rates \citep{Muratov2017, Mitchell2020}, and the chemical composition of the ambient CGM interacting with metal-rich outflows.} Through mixing, the metal-rich outflows are diluted, while the metal-poor CGM becomes relatively metal-rich \citep{FG2023}. This alteration in metallicities changes the gas cooling properties, thereby affecting the properties of gas accretion and wind recycling \citep{Hafen2020}, which can be crucial in understanding the galaxy stellar mass function \citep{Oppenheimer2010}.

\rev{The limited resolution in all simulations means that realistic astrophysical Reynolds numbers (Re) of order $\gtrsim 10^7$ are challenging to simulate \citep{Garnier2009}.} It is also not trivial to understand the effects of changing resolution -- \rev{both idealised \citep{ScannapiecoEA2015, SchneiderEA2017, Sparre2018} and cosmological simulations predict that the properties of the CGM change substantially with resolution. For instance, stellar mass and $\rm H\,\textsc{i}$ column densities change with resollution in cosmological simulations \citep{Murante2014, PakmorEA2016, Hopkins2018, vdv2019, Peeples2019, Hummels2019, Grand2021}}. Gas mixing is one fundamental process that changes with resolution \citep{Banda2018, vdv2021} \rev{because resolving the turbulent processes with better resolution reduces the numerical viscosity and diffusion, and models the cascade from larger to smaller scales of eddies better, effectively changing the turbulent viscosity \citep{landau2013fluid, Kemenov2011, Valdarnini2011, Schmidt2015}. Mesh-based codes usually rely on implicit mixing models rather than explicit models (for subgrid-scale mixing) used in particle-based codes (no mixing unless there is a subgrid model). Given enough resolution, one could identify the gas properties that cause mixing in codes with implicit mixing schemes and use them to calibrate explicit subgrid-scale mixing models. The difference in the numerical techniques changes the modelling of the turbulent cascade significantly \citep{Bauer2012}, and the need for different mixing models stems from the infeasibility to perform direct numerical simulations for astrophysical scenarios due to high Re \citep[computational effort $\propto$Re$^4$;][]{Meneveau2000}.}

\rev{There are several tricks for simulating the unresolved small-scale turbulent mixing. Large eddy simulations (LES) separate large and small scales using a spatial filter and subgrid models with an explicit turbulent viscosity or diffusivity for the unresolved gas mixing \citep{Leonard1975, Rogallo1984, Lesieur1996, Meneveau2000, Sagaut2006, Garnier2009, Schmidt2015}. The Smagorinsky model, which assumes that velocity shear fluctuations drive turbulent mixing, is a commonly adopted subgrid model due to its simplicity \citep{Smagorisnky1963, Schmidt2007, Shen2010, Shen2013, Brook2014, Williamson2016, Tremmel2017, Escala2018, Sokolowska2018}}\footnote{\rev{This is related to Boussinesq eddy viscosity hypothesis \citep{Schmidt2007}, Reynolds stress tensors \citep{Reynolds1895}, and Prandtl mixing length theory \citep{Prandtl1925, landau2013fluid}}.}. \rev{\citet{Wadsley2008} implemented the Smagorinsky model and another simple model based on velocity differences in a smoothed-particle hydrodynamics (SPH) code for kinetic and thermal energy dissipation, respectively, to obtain more accurate results. \rev{The inclusion of metal diffusion modelling in \citet{Shen2013} as opposed to \citet{Shen2012} increased the spread of heavy elements from outflowing wind material to the surroundings \citep{Shen2010}}. The Smagorinsky model, however, results in inaccuracies (e.g., overmixing, incompressibility, modelling supersonic flows due to the presence of shearing shocks), as it fails to describe a scale- and time-dependent diffusivity, because it was derived for highly non-magnetised subsonic flows without a bulk shear. As none of these assumptions hold in the ISM and CGM of galaxies, alternative models have been proposed and implemented recently.}

\rev{\citet{Colbrook2017} found superdiffusive mixing in supersonic turbulent ISM-like conditions; stressing the need to calibrate the Smagorinsky constant for non-turbulent, anisotropic, and non-shearing flows outside the inertial range of turbulence. They found that the effective diffusivity is proportional to the Mach number (also proportional to the local velocity dispersion) of the fluid \citep[see also ][]{Greif2009, Williamson2016}. Similarly, \citet{Rennehan2019} showed, with the help of isolated galaxy and cosmological simulations using Lagrangian techniques (SPH and meshless finite-mass methods), that the problems associated with the Smagorinsky model can be resolved if the Smagorinsky constant is a function of space and time \citep{Garmano1991}. \citet{Rennehan2021} highlighted how an explicit anisotropic eddy viscosity model for metal mixing in Lagrangian cosmological simulations impacts early galaxy evolution and alters metal distributions in circumgalactic gas. \citet{Huang2022} used a different subgrid model calibrated from higher-resolution cloud simulations and found a skewed but unimodal CGM metallicity distribution. The nature of mixing in the CGM radically alters the metallicity distributions; thus, it is crucial to understand the nature of mixing in realistic CGM-like conditions that include magnetic fields. Thus, we aim to validate the typical mixing scalings using our cosmological simulations with implicit mixing and magnetic fields, which are of great dynamical importance in galaxies.}

\rev{Two} main scenarios for the presence of magnetic fields are that large-scale outflows carry them from the ISM into the CGM \citep{Pakmor2020, Heesen2023, Ramesh2023, ArmburoGarca2023} or that a turbulent dynamo operates within the CGM \citep{Beck2012, Marinacci2015, Pakmor2017, Pakmor2020}. As a result, magnetic field strengths \rev{of $0.1-1 \muG$ }are observed up to impact parameters of a few tens of kpc from the galaxy disks into the CGM \citep{Bernet2013, LanP2020, AmaralEA2021, Shah2021, Bockmann2023}. \citet{vdv2021} showed the results of cosmological simulations of MW-mass galaxies with and without magnetic fields, which indicate that magnetic fields have a strong influence on CGM dynamics. On a fundamental level, magnetic fields can directly affect mixing in the CGM by \rev{suppressing} Kelvin-Helmholtz instabilities, \rev{leading to \rev{a reduced formation of} cool gas from the lack of intermediate temperature gas \citep{chandrasekhar1968hydrodynamic, Ji2019, Das2023, Jennings2023}}. In general, magnetic fields have been shown to reduce the required turbulent pressure support by providing magnetic pressure support \citep{Mohapatra2021b, Mohapatra2021a}, \rev{thus} suppressing turbulent mixing. \rev{Therefore,} including magnetic fields, as we do in this work, is important for studying the gas mixing properties of the CGM \citep{Pakmor2020, vdv2021}.

In this work, we use cosmological ‘zoom-in’ simulations of a single MW-mass halo with enhanced resolution in the CGM \citep{vdv2021}. We study the mixing of tracer dyes to understand the dependence of gas mixing on the physical properties of the CGM. The simulation method and the selection of tracer dye locations are described in \autoref{sec:methods}. In \autoref{sec:results}, we present the results showcasing the correlation of gas mixing with various gas properties, effects of limited resolution, and turbulent properties of the CGM. We summarise our findings in \autoref{sec:conclusions}.

\section{Methods}
\label{sec:methods}
\subsection{Simulations}
\begin{figure*}
    \includegraphics[width=1.75\columnwidth]{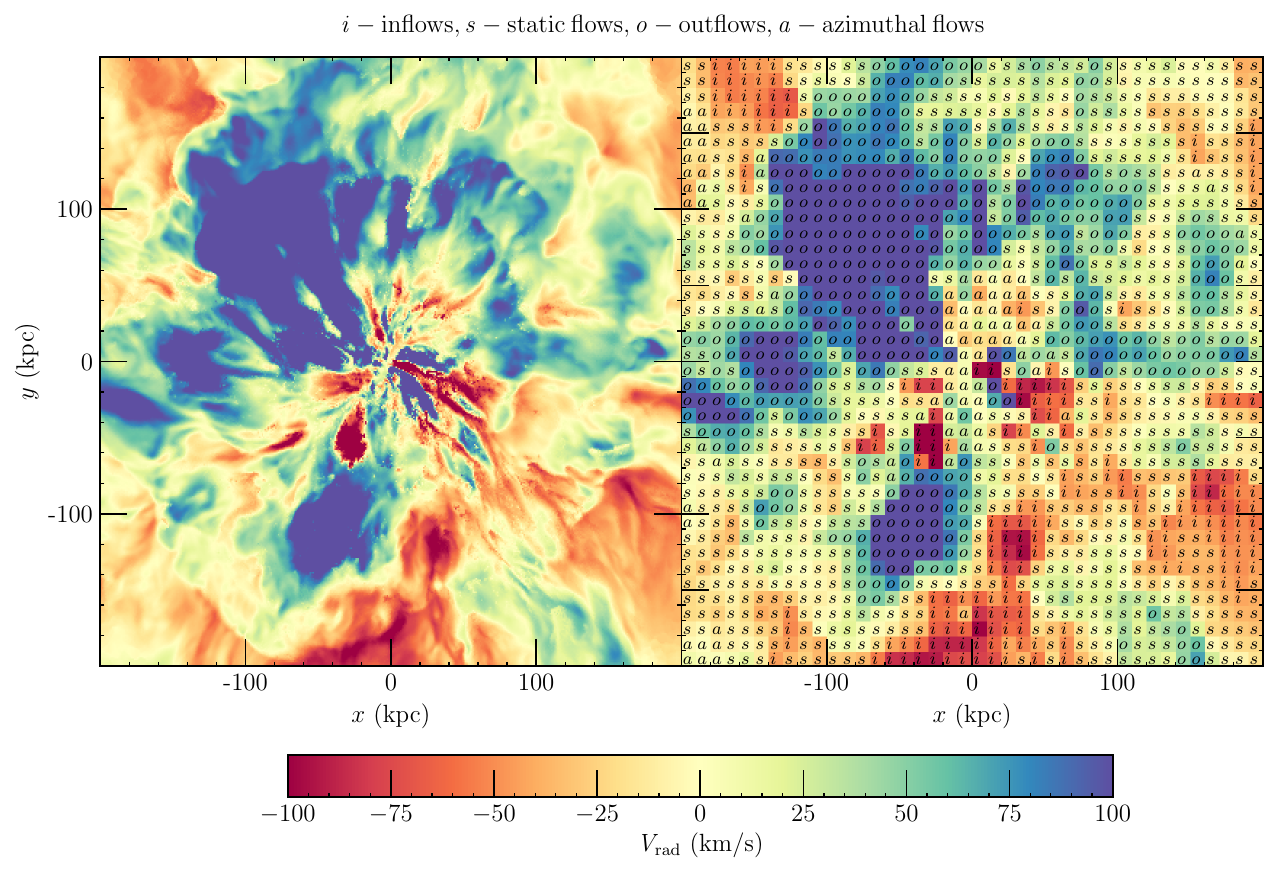} 
    \caption{{\it Left}: $10 \kpc$ deep (along the line-of-sight) projection through the halo centre of the radial velocity with standard mass resolution + $1 \kpc$ refinement in a box of $400 \times 400 \times 10 \kpc$. {\it Right}: Radial velocity of the \rev{voxelated (voxel} size $=10 \times 10 \times 10 \kpc$) CGM in a $400 \times 400 \kpc$ image. The text in the centre of every \rev{voxel} shows the flow type (see title) of the \rev{voxel} based on its radial and total velocities (see \autoref{sec:CGM_classification}). The galaxy is shown in the (random) coordinate frame of \rev{voxel} selection, not edge-on as in subsequent figures. }
    \label{fig:binned_h12_class}
\end{figure*}
This work utilises simulations from the Simulating the Universe with Refined Galaxy Environments (SURGE) project \citep[see also][]{vdv2019}, using the same initial conditions and galaxy formation model as the Auriga simulations, a suite of cosmological zoom-in simulations of MW-like galaxies \citep{GrandEA2017, vdv2021}. They were run with the \textsc{arepo} code, which employs moving-mesh techniques based on a Voronoi mesh \citep{Springel2010, PakmorEA2016, WeinbergerEA2020}. This quasi-Lagrangian scheme offers several benefits over other commonly used Eulerian (fixed mesh) codes with or without adaptive mesh refinement (AMR) and \rev{SPH} codes. It improves upon suppressed fluid instabilities in SPH \citep{SijackiEA2012} and Galilean non-invariance and the presence of overmixing in Eulerian codes \citep{Springel2010}. Additionally, several benefits of SPH (e.g. low numerical diffusion) and Eulerian (e.g. good convergence for smooth flows, capturing shocks accurately) codes are retained in these moving-mesh techniques \citep{Dale2015, PakmorEA2016}. Thus, moving mesh techniques should provide a more accurate physical description of gas mixing in the CGM \citep{DuffellEA2011}. 

For this study, we selected one of the Auriga galaxies (halo 12)\footnote{Information on the basic properties of simulated haloes with standard resolution is available at \url{https://wwwmpa.mpa-garching.mpg.de/auriga/about.html}. Details of halo 12 can be found in \citet{GrandEA2017} and \citet{vdv2021}} and resimulated it using standard mass refinement only \citep{vdv2021} and using standard mass refinement combined with fixed additional spatial refinements ($1 \kpc$ and $2 \kpc$). For all simulations, there is a standard mass refinement such that the target cell masses for baryons and dark matter particles are $5.4 \times 10^4 \Msun$ and $2.9 \times 10^5 \Msun$, respectively. For the runs with minimum $1$ and $2\kpc$ spatial resolutions, an additional volume refinement criterion is included within $1.2 \Rvir$, which is dominant when $\rho<\rho_{\rm threshold}/64$ and $\rho<\rho_{\rm threshold}/512$\footnote{$\rho_{\rm threshold}$ is the gas density above which star formation occurs, i.e., $\nH^{\star}=0.11 \cm^{-3}$.}, respectively. This translates to a maximum cell size of $\sim 1$ and $2 \kpc$ and a maximum cell volume of $\sim 1$ and $8 \kpc ^3$ inside $1.2 \Rvir$, respectively. Here, we define the virial radius ($\Rvir$) as the radius within which the mean overdensity ($\rhoover$) is 200 at its redshift, where $\rhoover$ is the density divided by the critical density of the universe. Mass refinement is dominant when $\rho>\rho_{\rm threshold}/64$ and $\rho>\rho_{\rm threshold}/512$, leading to spatial resolutions $\le 1$ and $2\kpc$ overall, respectively. The majority of our results use $1 \kpc$ resolution runs, and the $2 \kpc$ runs are used \rev{only} for \rev{the} convergence tests and checking the results' robustness (see \autoref{Sec:num_mix}). The selected halo has $\Rvir \approx 213 \kpc$ at $z \approx 0.059$ (subsequent analyses performed at or after this redshift), and the total \rev{virial} mass of the main halo is $10^{12} \Msun$.

The simulations were run with the latest version of \textsc{arepo} while adopting a $\rm \Lambda$CDM cosmology with parameters taken from \citet{Planck2014}: $\rm \Omega_m = 1-\Omega_{\Lambda} = 0.307$, $\rm \Omega_b = 0.048$, $h = 0.6777$, $\sigma_8 = 0.8288$, and $n = 0.9611$. Except for the additional spatial refinement, we use the same physical model as the original Auriga simulations \citep{GrandEA2017}. However, mass and metals ejected by a star particle are injected only into its host cell rather than its 64 neighbouring cells as in \citet{GrandEA2017}. This is not expected to bring significant differences within the scope of this study \rev{because metals quickly mix with surrounding gas and there are numerous such metal injection events (supernovae) that both methods would yield statistically similar results \citep[see][who studied the difference in the context of rare rapid neutron capture elements and found negligible impact]{vandevoort2020}. It should be noted that momentum injection is still performed in the same way as the original Auriga simulations with wind particles, so outflow properties are not altered. Furthermore, the gas we study is so far away in the CGM that the precise details of the metal injection methodology would not affect the CGM properties.}

\subsection{Dye Location Selection}
\label{sec:dye_location_selection}
To understand gas mixing in different environments of the CGM, we implement tracer dyes (referred to as `dye' hereon) in our simulations. The dye is a passive scalar that advects with the fluid. To initialise the dye, we select cells according to the criteria given in \autoref{sec:dye_location} at $z \approx 0.059$, inject the dye, and restart the simulation from that point with outputs saved every $10 \Myr$. The selected cells for dye injection are assigned with a scalar value of 1, the rest being 0. At any later time in a given cell, the mass fraction of the initially injected dye can be determined by the scalar value in that cell. Thus, it allows us to track and analyse gas transport between the computational cells. Previously, a similar methodology has been adopted in other simulations with different terminology, for instance, pollutants in \citet{PanEA2012}, passive tracers in \citet{Gronnow2018}, and passive scalars in \citet{SchneiderEA2020}. This dye allows us to track the entire distribution of mixed gas, useful for studying advection and diffusion processes.

\subsubsection{CGM classification}\label{sec:CGM_classification}

To capture the multiphase nature of the CGM, we inject the dye in 95 cells at locations representing different local environments with different physical properties at $z=0.059$ and create outputs with $10 \Myr$ spacing for a total of $200 \Myr$. \rev{First, we average gas properties into a uniform 3D grid, centred on the galaxy, with dimensions $400 \times 400 \times 400$ kpc. The 3D voxels are $10 \times 10 \times 10$ kpc cubes that inherit averaged gas properties from the gas cells within each region.} Each ($10 \kpc$)$^3$ \rev{voxel} contains $\sim 10^3$ cells as the individual cell sizes are $\sim 1 \kpc$; thus, $10 \kpc$ \rev{voxel} gas properties are determined using either averages or the sum of different gas properties of cells inside the \rev{voxel}. The volume-weighted average is used for temperatures, densities, and metallicities, the mass-weighted average is used for radial velocities and velocities, and the sum is used for kinetic energy.

All the \rev{voxels} with overdensities ($\rhoover$\,) $\geq 150$ are ignored in selection to avoid choosing the ISM \rev{or disk-halo interface} gas of either the host or the satellite galaxy\rev{, and make the selection sufficiently far into the CGM ($\gtrsim 20 \kpc$ from the galactic centre).} \rev{Voxelating} prevents identifying different environments based on fluctuations at individual cell scales and selects regions based on their $\rm 10\,kpc$ local scale properties. All the gas properties concerning the \rev{voxels} are represented with a \rev{`$v$'} in the superscript. For instance, the radial (total) velocity of an individual \textsc{arepo} cell is denoted as $\Vrad$ ($V$), while that for the interpolated \rev{voxel} is denoted $\pVrad$ ($\pV$).

\autoref{fig:binned_h12_class} (left panel) shows a thin projection of the radial velocities with strong outflows in blue and inflows in red. The right panel is the \rev{voxelisation} of the left panel. We further characterise the \rev{voxels} into inflows, outflows, static, and azimuthal flows locally using the following criteria. We define $|\pVrad| > 40 \kms$ as either an inflow ($\vec{\pVrad}$ is negative) or an outflow ($\vec{\pVrad}$ is positive). And when $|\pVrad| < 40 \kms$, it is defined as either a static flow ($|\pV| < 80 \kms$) or an azimuthal flow ($|\pV| > 80 \kms$). Static flows are relatively static regions. \rev{The net inflow and outflow velocities have magnitudes of $\sim 50 \kms$ in the presence of magnetic fields in the CGM \citep[see Fig. 6 in][]{vdv2021}. As outflows would slow down due to entrainment as they reach the outer CGM, we select a $40 \kms$ radial velocity cutoff. The total velocity cutoff of $80 \kms$ is chosen to differentiate whether the non-radial gas has a strong component in the azimuthal direction.} 

\subsubsection{Dye injection location based on flow interactions and gas properties}
\label{sec:dye_location}
We identify locations for dye injection based on flow interactions: interacting inflows-outflows ($io$), pure inflows ($i$), pure outflows ($o$), pure static flows ($s$), and outflows interacting with static flows ($os$). 
\begin{itemize}
    \item Interacting inflows-outflows ($io$): To find these, we examine \rev{voxels} representing inflow and outflow regions nearby, specifically within a distance equal to twice the \rev{voxel} diagonal ($20 \sqrt{3} \kpc$). Assuming constant velocities, we track their trajectories to determine the points of the closest approach. We ensure that this occurs within a reasonable timeframe ($\sim 150 \Myr$). The interaction is confirmed if both \rev{voxels} reach their closest points nearly simultaneously. This method ensures that only plausible inflow-outflow interactions are considered. We select both the inflow and outflow \rev{voxels} from an inflow-outflow pair but dye them with separate colours. Different dye colours refer to different initialized passive scalars, allowing them to track the dye independently of one another. More details can be found in \autoref{sec:appendix_io}.    
    
    \item Pure inflows ($i$), outflows ($o$), and static flows ($s$): If more than $90\%$ of the \rev{voxels} in a region of $50 \times 50 \times 50 \kpc$ are classified as a single flow type (see \autoref{sec:CGM_classification}) along with the central \rev{voxel}, then the central \rev{voxel} is selected as a pure inflow, outflow, or a static flow, depending on the majority of the flow type in that region.

    \item Outflows interacting with static flows ($os$): To select static flows with outflow interaction at a later time, we first find \rev{voxels} with strong outflows ($\Vrad>100 \kms$). Then we select regions of $50 \times 50 \times 50 \kpc$ around the strong outflows with static flows $\ge 30$ per cent, outflows $\ge 50$ per cent, and inflows $\le 5$ per cent, to locate regions with substantial outflows and static flows. The static \rev{voxel} whose position vector (from the centre of the strong outflow) has the smallest angle with the strong outflow velocity vector inside a region of $50 \times 50 \times 50 \kpc$ is selected as $os$. 
    
\end{itemize}

\begin{figure*}
\begin{subfigure}{1\textwidth}
\centering
\includegraphics[width=1\columnwidth]{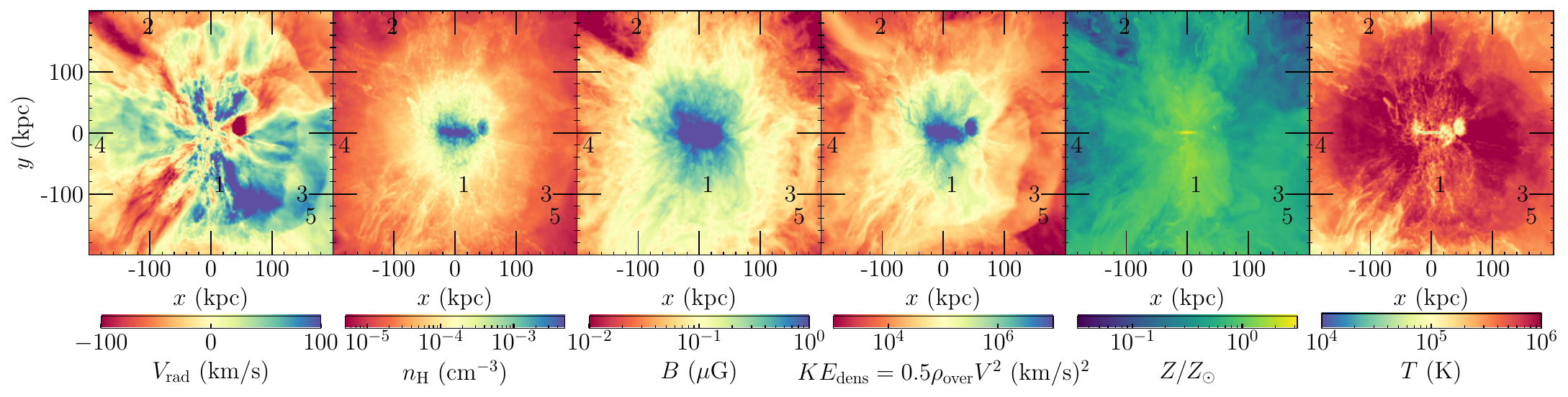} 
\end{subfigure}
\begin{subfigure}{1\textwidth}
\includegraphics[width=1\columnwidth]{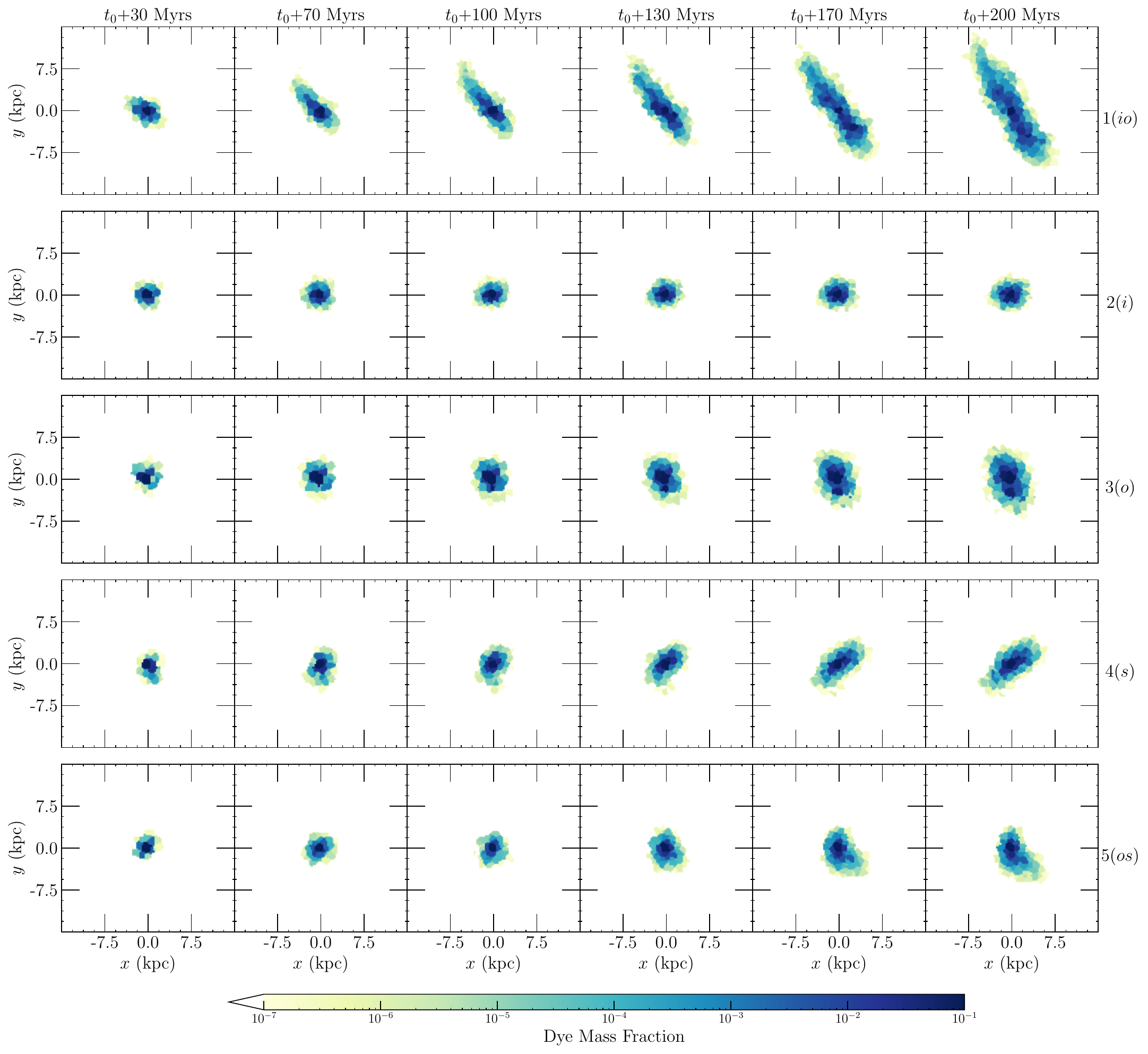} 
\end{subfigure}    
\caption{Top panels show edge-on (rotated such that the stellar disc is edge-on) gas projections centred on the main galaxy with $100 \kpc$ depth and $400 \kpc \times 400 \kpc$ width. From left to right, the panels show the radial velocity, hydrogen number density, magnetic field strength, kinetic energy density, metallicity, and temperature. The five numbers in the \rev{top panels} show some locations of dye injection. The evolution of dye at the five locations over $200 \Myr$ is shown in the bottom panels in a $40 \times 40 \kpc$ box. $\rm t_0$ is the time the dye was injected into single cells (with a dye mass fraction of 1). Here, the row labeled $1(io)$ is an outflow from interacting inflows-outflows ($io$), $2(i)$ is a pure inflow ($i$), $3(o)$ is a pure outflow ($o$), $4(s)$ is a pure static flow ($s$), and $5(os)$ is an outflow interacting with a static flow ($os$). }
\label{fig:flashyfig}
\end{figure*}

\begin{figure*}
    \includegraphics[width=2\columnwidth]{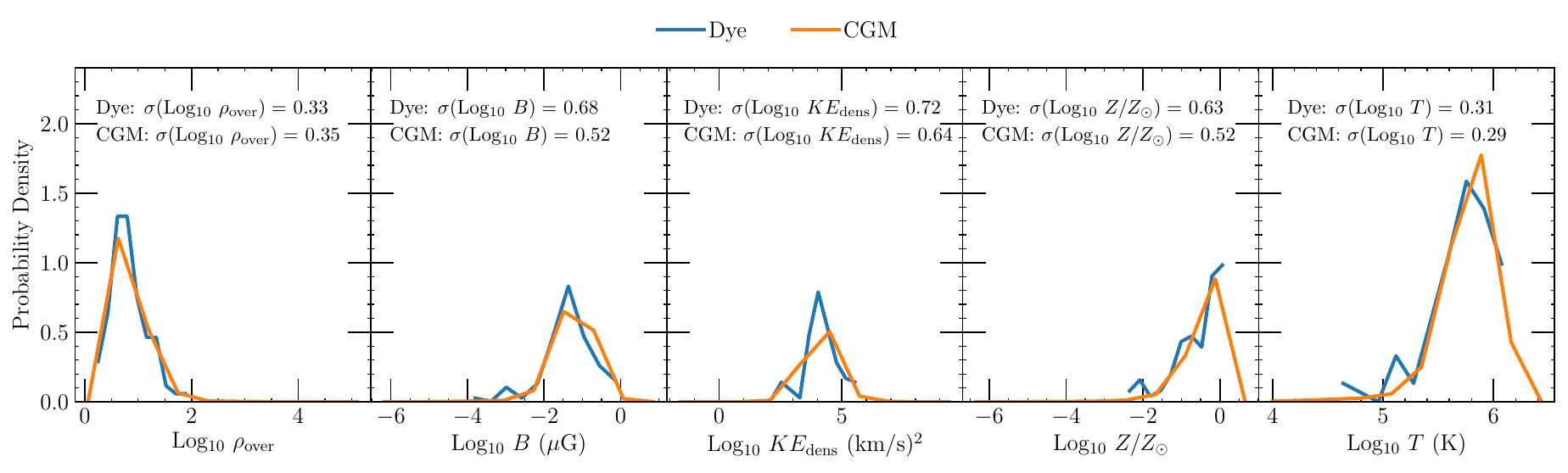} 
    \caption{The probability density function of several volume-weighted gas properties of the CGM (orange) and the dye injection cells (blue) with logarithmic binning (bins=10). The text in every panel shows the standard deviation of the gas property values (in log$_{10}$) for the dyed cells and the CGM ($50 \kpc<r<200 \kpc$). The standard deviation for the dyed cells is similar to that of the CGM for all gas properties, meaning that the dye selection is from a diverse range of environments that represent the CGM.}
    \label{fig:dye_stats}
\end{figure*}

\begin{table*}
\caption{Properties and statistics of the dye injection locations in the CGM. For gas properties, `L' refers to low, `H' refers to high, and `N/A' refers to a random selection. The columns are as follows: 1: number of the dye colour, 2: type of flow interaction, 3: number of separate dye injection cells, 4: temperature selection, 5: kinetic energy selection, 6: metallicity selection. In total, 95 different cells are selected.}
\begin{tabular}{|c|c|c|c|c|c|}

\hline
$N_{\rm col}$ & $\rm Flow~interactions$ & $\rm No.~of~dye~injection~cells$ & $T$ & $KE$ & $Z/Z_{\odot}$ \\
\hline
0 & $\rm Inflows~in~interacting~inflows-outflows~(\mathit{io})$ & 4 & L & L & L \\
1 & $\rm Inflows~in~interacting~inflows-outflows~(\mathit{io})$ & 2 & L & H & H \\
2 & $\rm Inflows~in~interacting~inflows-outflows~(\mathit{io})$ & 2 & L & L & L \\
3 & $\rm Inflows~in~interacting~inflows-outflows~(\mathit{io})$ & 4 & H & L & L \\
4 & $\rm Inflows~in~interacting~inflows-outflows~(\mathit{io})$ & 4 & H & H & H \\
5 & $\rm Outflows~in~interacting~inflows-outflows~(\mathit{io})$ & 4 & H & L & L \\
6 & $\rm Outflows~in~interacting~inflows-outflows~(\mathit{io})$ & 2 & H & H & H \\
7 & $\rm Outflows~in~interacting~inflows-outflows~(\mathit{io})$ & 2 & L & L & L \\
8 & $\rm Outflows~in~interacting~inflows-outflows~(\mathit{io})$ & 4 & H & L & L \\
9 & $\rm Outflows~in~interacting~inflows-outflows~(\mathit{io})$ & 4 & H & H & H \\
10 & $\rm Pure~inflows~(\mathit{i})$ & 4 & L & L & L \\
11 & $\rm Pure~inflows~(\mathit{i})$ & 4 & L & H & H \\
12 & $\rm Pure~outflows~(\mathit{o})$ & 4 & H & L & L \\
13 & $\rm Pure~outflows~(\mathit{o})$ & 3 & H & H & H \\
14 & $\rm Pure~static~flows~(\mathit{s})$ & 4 & L & L & L \\
15 & $\rm Pure~static~flows~(\mathit{s})$ & 4 & L & L & H \\
16 & $\rm Pure~static~flows~(\mathit{s})$ & 4 & L & H & L \\
17 & $\rm Pure~static~flows~(\mathit{s})$ & 4 & L & H & H \\
18 & $\rm Pure~static~flows~(\mathit{s})$ & 4 & H & H & L \\
19 & $\rm Pure~static~flows~(\mathit{s})$ & 4 & H & H & H \\
20-25 & $\rm Outflows~interacting~with~static~flows~(\mathit{os})$ & 24 & N/A & N/A & N/A \\
\end{tabular}
\label{tab:init_table}
\end{table*}

Finally, we select an individual cell (i.e., one resolution element) inside each \rev{voxel} ($10 \times 10 \times 10 \kpc$ region) to inject the dye. We only consider individual cells with gas properties ($T$, $Z$, and $KE$) in similar ranges (low or high) as their respective \rev{voxels}. For interacting inflows-outflows, the cells (one from inflow and one from outflow \rev{voxel}) closest to each other are selected. For pure flows, the cell closest to the centre of the \rev{voxel} is selected. For outflows interacting with static flows, the cell (from the static flow \rev{voxel}) closest to and in the direction of the strong outflow is selected. The dye injection cells in a specific colour are separated by at least $60 \kpc$ to ensure no overlap. \autoref{tab:init_table} shows the dyed cells' gas properties and flow interaction types based on our selection criteria. No. of dye injection cells refers to the number of cells dyed with one colour. `L' (bottom $33$ percentile) and `H' (top $33$ percentile) refer to low and high values for those gas properties. `N/A' refers to a random selection because the sample size was not large enough to subdivide into different gas properties. In total, 95 cells are dyed with 26 different dye colours. The procedures described in this section ensure a sample with diverse flow interactions and varying gas properties.

\autoref{fig:flashyfig} shows five dyed cells, with the top panels showing their locations and different gas properties at the injection time. Numbers 1-5 (in the top \rev{panels}) represent one dye injection cell from each flow interaction type - 1 interacting inflow-outflow, 2 pure inflow, 3 pure outflow, 4 pure static flow, and 5 outflow interacting with static flow. The bottom panels show the evolution of dye mass fraction over $200 \Myr$ at six different times. The dye mixes more in the interacting inflow-outflow case (more shearing) while mixing is relatively slower in pure (more coherent) flows. Thus, there could be some dependence of dye mixing on the gas velocity statistics.

\autoref{fig:dye_stats} compares the probability density functions (PDF) of gas properties of the dyed cells (95 in number) and overall CGM gas (cells between $50<r<200 \kpc$). Both PDFs look very similar for all gas properties. The dyed cells' standard deviations (given in each panel) are usually somewhat higher than or almost equal to the standard deviations of the CGM gas. This shows that the 95 cells selected for dye injection represent a diverse environment of the CGM in terms of various gas properties like density, temperature, magnetic fields, kinetic energy density, and metallicity. Note that we represent gas densities using two quantities throughout the text - hydrogen number density ($\nH$) and overdensity ($\rhoover$). Multiplying $\nH$ with a factor of $2.44 \times 10^5$ approximately converts it to $\rhoover$.

\section{Correlating gas mixing with physical properties}
\label{sec:results}
\begin{figure*}
    \includegraphics[width=2\columnwidth]{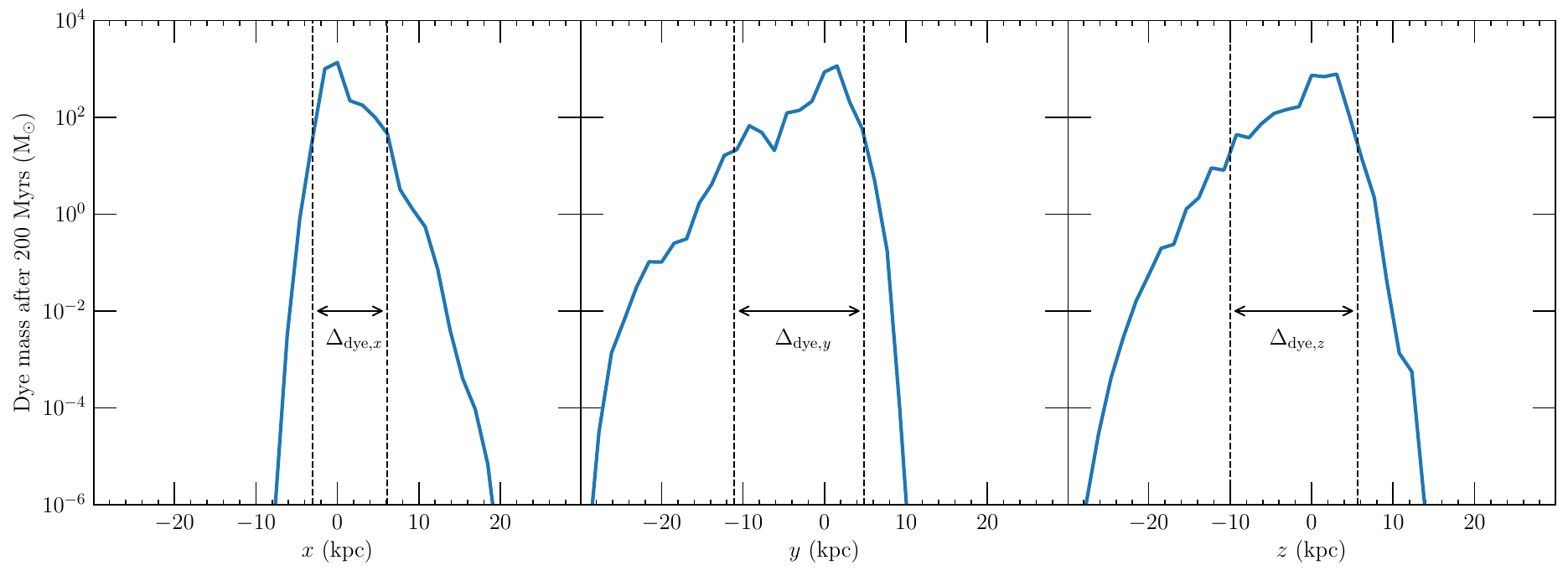} 
    \caption{Dye mass $200 \Myr$ after dye injection along the three coordinate axes in a single dye region. The dashed lines represent the $\rm 1^{st}$ and $\rm 99^{th}$ dye mass percentile. The width of the dashed lines ($\spreadx, \spready, \spreadz$) is the dye spread in the $x$, $y$, and $z$ directions, respectively. }
    \label{fig:spread_quant}
\end{figure*}

To understand gas mixing in the CGM, we first quantify how much the dye mixes and spreads in its environment. We study the surroundings (up to $60 \kpc$) of the dye injection region, centred on the cell with the largest dye mass fraction. The $60 \times 60 \times 60 \kpc$ region is divided into 40 $1.5 \times 60 \times 60 \kpc$ bins along coordinate axes $x, y, z$. The dye mass in each bin is added and shown in \autoref{fig:spread_quant} along the three coordinate axes. Cells with dye mass fraction $\le 10^{-7}$ are excluded from this calculation due to a small fraction of residual dye spreading because of numerical effects. We define the dye spread, $\spread$, in any direction as the width from the $\rm 1^{st}-99^{th}$ percentile of the initially injected dye mass. Thus, $\spreadx$, $\spready$, and $\spreadz$ are the sizes of the cloud of dyed material along the three coordinate axes (as shown in \autoref{fig:spread_quant}). We try other definitions for the spread like standard deviations and full width at half maximum (FWHM) with the spread values reducing $\approx$ ten-fold and two-fold, respectively. \rev{Most} of our results and conclusions remain unaffected by the choice of definition of spread. \rev{We use $\spread$ from $\rm 1^{st}-99^{th}$ percentile definition everywhere in the paper except \autoref{sec:diffusion_coefficient} where we use the standard deviation definition, which is more appropriate for diffusion coefficient calculations. The reason for choosing $\rm 1^{st}-99^{th}$ percentile definition everywhere else is because it probes most of the dye mass, representing the true shape of the dye cloud more accurately.} 

When discussing the dye spread below, we generally refer to the size of the dyed cloud after $200 \Myr$ of evolution, unless otherwise stated (usually where we study the temporal evolution of dye spread, e.g., \autoref{Sec:num_mix}). For reference, the eddy turnover time for a $5 \kpc$ region in this simulation is $\sim 200 \Myr$ \rev{($l/\vdis \sim 5\kpc/25 \kms$)}, the CGM dynamical and cooling time scales are typically in the order of magnitude of $\sim 1 \Gyr$, and the free-fall timescales are a few hundred $\Myr$ \citep{TumlinsonEA2017, Stern2019, Hafen2020}. So, $200 \Myr$ should be sufficient to look at turbulent mixing without the character of the CGM changing significantly. It should be noted that the \rev{the dye both advects and diffuses}. So, we take the cell with the maximum dye fraction as the centre. The resultant moving boxes' average velocity direction is aligned with the bulk velocity direction (averaged over $10 \kpc$) at the dye injection time. Thus, bulk velocities provide a good estimate for determining the translation/advection of the dye.

\subsection{Magnitude of Gas Mixing}
\label{sec:mag_gas_mixing}
\begin{figure*}
    \includegraphics[width=2\columnwidth]{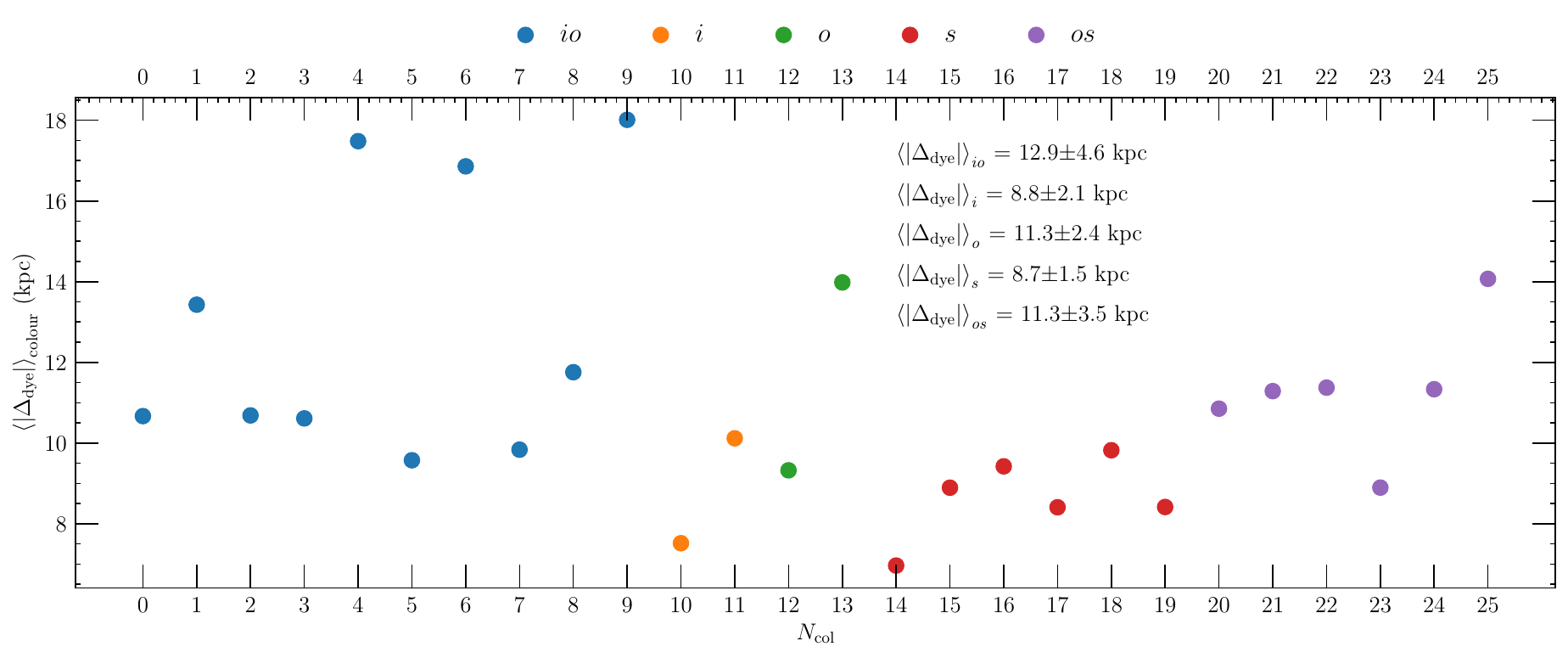} 
    \caption{The size of the dyed area after $200 \Myr$ ($|\spread|$) averaged over injection regions with the same dye colours (see \autoref{tab:init_table}). Different coloured points in the graph depict different types of gas flow interactions ($io$, $i$, $o$, $s$, $os$), and the text inset lists the average of their $|\spread|$. $io$ are interacting inflow-outflow pairs, $i$ are pure inflows, $o$ are pure outflows, $s$ are pure static flows, and $os$ are outflows interacting with static flows. The inflow-outflow pairs ($io$) exhibit the highest dye spread, while pure inflows ($i$) and static flows ($s$) are among the lowest. }
    \label{fig:spread_colour}
\end{figure*}

To quantify gas mixing, we use the magnitude of dye spread, i.e. the size of the area containing significant amounts of dye, as $|\spread| = (\spreadx^2 + \spready^2 + \spreadz^2)^{1/2}$. Note that the choice of the axes in the coordinate system is arbitrary and there is no preferential dye spread along any of the coordinate axes. We group dyed regions together that have similar physical properties (see \autoref{tab:init_table}) and refer to each group as a different `colour'. First, we explore differences in $|\spread|$ for different colours corresponding to different flow interactions and physical properties. Since dye spread is caused by the gas mixing with its surroundings, we use the terms gas mixing and dye spread interchangeably from hereon.

\autoref{fig:spread_colour} shows the spread averaged over dye regions with similar properties (see \autoref{tab:init_table}). $|\spread|$ averaged over different flow interactions ($io$, $i$, $o$, $s$, $os$) regardless of their $T$, $KE$, or $Z$ is shown as text in \autoref{fig:spread_colour}. $|\spread|$ for interacting inflows-outflows (blue, $\langle |\spread| \rangle_{io}$) is on average (but with large scatter) greater than other flow interactions. The spread for the pure flows ($i$, $o$, $s$) is generally lower than locations with interacting flows ($io$, $os$). Since pure flows are more coherent, their motion is relatively ordered on $\mathcal O$($10 \kpc$) scales, whereas interacting flows are expected to create more turbulence and have a higher velocity dispersion.

For each flow interaction without the static gas, colour numbers 1, 4, 6, 9, 11, and 13 have a higher $|\spread|$ than other colours in the corresponding flow interactions. From \autoref{tab:init_table}, it is clear that these colour numbers correspond to regions with higher kinetic energy. We do not see such a correlation with other selected properties - temperature and metallicity. However, this dependence could be indirect, as gas properties are often correlated in CGM environments. Therefore, we will quantify the correlation between $|\spread|$ and various local gas properties \rev{like temperature, metallicity, density, kinetic energy, and magnetic fields, and derived gas properties like velocity dispersion, and partial derivatives of the velocity field}

\begin{figure*}
    \includegraphics[width=2\columnwidth]{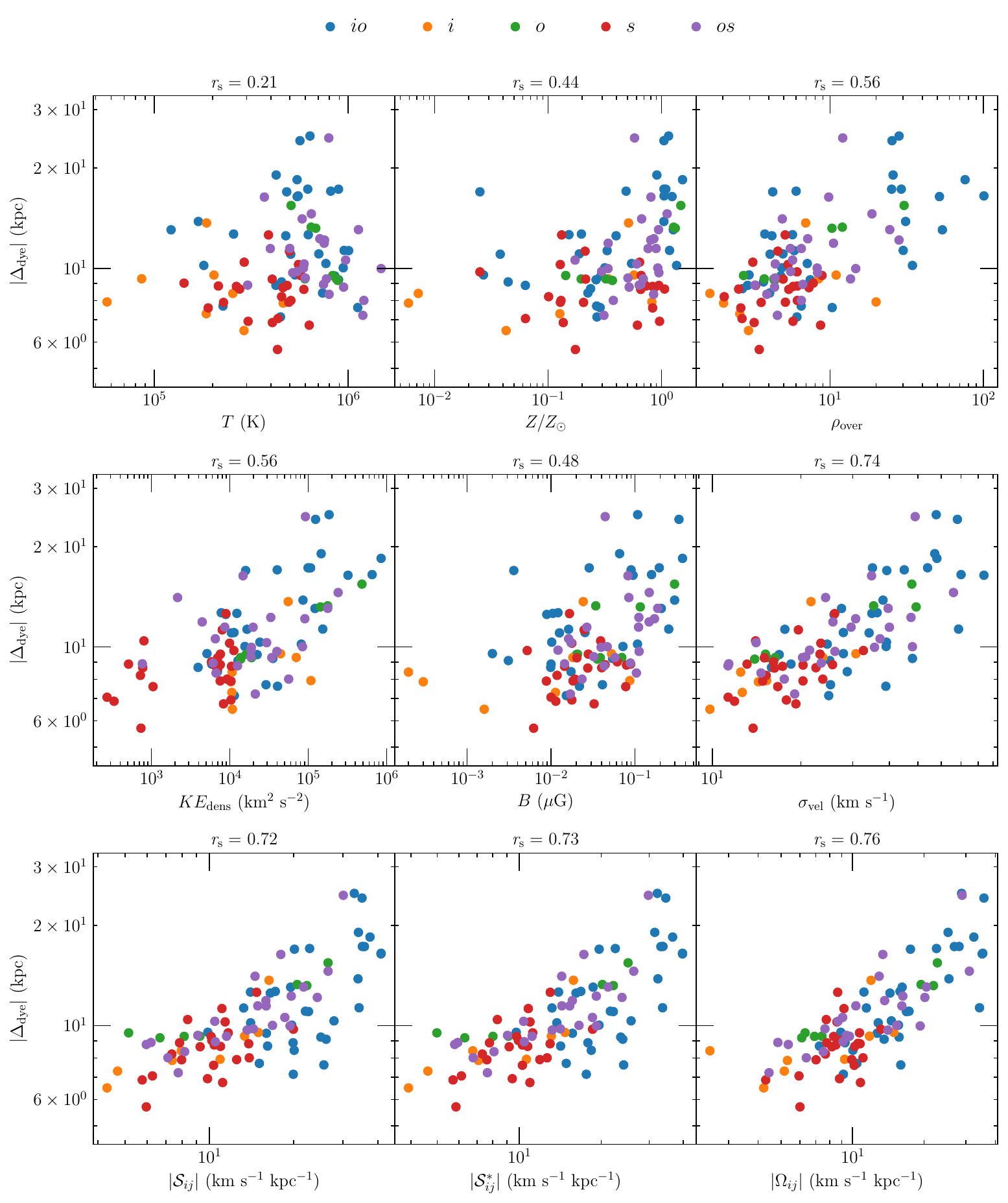} 
    \caption{The magnitude of dye spread after $200 \Myr$ as a function of different gas properties averaged in a $10 \kpc$ region around the dye injection locations at injection time. The Spearman rank correlation coefficient ($\rs$) is shown in the panel titles. \rev{From left to right, the top panels show tracer spread correlations with temperature, metallicity, and overdensity, the middle panels show correlations with kinetic energy density, magnetic field, and velocity dispersion, and the bottom panels show the correlations with symmetric, traceless symmetric, and antisymmetric parts of the shear tensor (see \autoref{eq:T_ij}). Both velocity dispersion and tensor quantities correlate strongly with dye spread.} The kinetic energy density ($\KEdens$) and overdensity ($\rhoover$) also show a significant correlation, possibly due to their dependence on $\vdis$ (see \autoref{tab:init_corr_table}). \rev{P-values for gas properties from left to right in the top panel are $4.4 \times 10^{-2}$, $8.7 \times 10^{-6}$, $4 \times 10^{-9}$; in the middle panel, $3.3 \times 10^{-9}$, $1 \times 10^{-6}$, $1.3 \times 10^{-17}$; and in the bottom panel, $4.1 \times 10^{-16}$, $5.1 \times 10^{-17}$, and $3.1 \times 10^{-19}$.}}
    \label{fig:spread_gasprops}
\end{figure*}

\rev{\autoref{fig:spread_gasprops} shows (for all 95 selected dye injection locations) the correlation of $|\spread|$ with several gas properties averaged within the $10 \times 10 \times 10 \kpc$ region centred on the dye injection location at injection time. Here, kinetic energy density is the volume-specific kinetic energy, i.e., $\KEdens = 0.5 \rhoover |\vec{V}|^2$, and velocity dispersion ($\vdis$) is the root mean square of the standard deviation of the velocity vectors ($\vec{V}$). For the local gas properties, we use volume-weighted average temperature ($T$), metallicity ($Z/Z_{\odot}$), overdensity ($\rhoover$), and magnetic field ($B$). The top and middle panels show the correlations of the properties mentioned above. The bottom panel shows the correlations with symmetric ($\mathcal{S}_{ij}$), traceless symmetric ($\mathcal{S}_{ij}^*$), and antisymmetric shear tensors ($\Omega_{ij}$). All of these are derived from the shear tensor,}

\begin{ceqn}
\begin{align}\label{eq:T_ij}
    \mathcal{T}_{ij} = \frac{\partial V_i}{\partial x_j},
\end{align}
\end{ceqn}
\rev{made up of the partial derivatives of the velocity field. The symmetric part of this tensor $\mathcal{T}_{ij}$ is}

 \begin{ceqn}
\begin{align}\label{eq:S_ij}
    \mathcal{S}_{ij} = \frac{1}{2}(\mathcal{T}_{ij} + \mathcal{T}_{ji}),
\end{align}
\end{ceqn}
\rev{the traceless symmetric part of the tensor $\mathcal{T}_{ij}$ is}

\begin{ceqn}
\begin{align}\label{eq:S_ij*}
    \mathcal{S}_{ij}^* = \frac{1}{2}(\mathcal{T}_{ij} + \mathcal{T}_{ji})-\frac{1}{3}\delta_{ij}\mathcal{T}_{kk},
\end{align}
\end{ceqn}
\rev{where $\delta_{ij}$ is the Kronecker delta tensor. The antisymmetric part of the shear tensor is}

\begin{ceqn}
\begin{align}\label{eq:O_ij}
    \Omega_{ij} = \frac{1}{2}(\mathcal{T}_{ij} - \mathcal{T}_{ji}).
\end{align}
\end{ceqn}

\rev{While $\mathcal{S}_{ij}$ represents the strain rate (deformation) of the fluid, its traceless component $\mathcal{S}_{ij}^*$ represents pure shear deformation without any volume change. $\Omega_{ij}$ captures the local rigid-body rotation (vorticity) of the fluid}\rev{. Note that we will often explore the correlations with three tensors, i.e. symmetric shear tensor $\mathcal{S}_{ij}$, traceless symmetric shear tensor $\mathcal{S}_{ij}^*$, and antisymmetric shear tensor $\Omega_{ij}$. Thus, the term "tensor quantities" in subsequent parts of the study refers to these three tensors in general, and the term "velocity-derived" quantities refers to $\vdis$ and the tensor quantities \citep[check][for an overview of the tensors]{beattie2025}.}

To quantify the correlation of dye spread with gas properties, we perform the Spearman rank correlation test, which determines the strength and direction of the monotonic relationship between any two real variables. The Spearman coefficient ($\rs$) varies between -1 and 1, where -1 indicates a strong negative relationship, 0 indicates no relationship, and 1 indicates a strong positive relationship. \rev{All the p-values (mentioned in the caption of \autoref{fig:spread_gasprops}) for the null hypothesis that the data sets in are uncorrelated are $< 10^{-5}$, except for temperature ($\sim 0.05$), which only shows a weak correlation.} \autoref{fig:spread_gasprops} shows $\rs$ values above each panel; there is some correlation between dye spread and \rev{the magnitudes of} all the gas properties\footnote{\rev{The magnitude of any tensor $\mathcal{T}_{ij}$ is derived as $|\mathcal{T}_{ij}| = (2\mathcal{T}_{ij}*\mathcal{T}_{ij})^{1/2}$, also known as the “Frobenius norm” of a matrix \citep[same as][]{Piomelli1995}, and is coordinate independent.}}. Some (such as overdensity, kinetic energy density, velocity dispersion\rev{, and tensor quantities}) have reasonably high $\rs$ values $\geq 0.5$. This multivariate dependence is because CGM properties depend on one another. \autoref{tab:init_corr_table} shows the results of the Spearman rank correlation test between gas properties, some of which have high $\rs$ values of $\geq 0.6$. In the CGM, properties like $\rho$ decrease radially due to gravity-dominated structure formation, but $B$, and $Z$ decrease because they are enhanced in the central galaxy and then distributed by feedback \citep{vdv2021}. Additionally, compressing the gas will increase its density and $B$, while expanding will do the opposite. Velocity dispersion ($\vdis$) is usually high where CGM flows interact at high velocities, leading to high $KE$. Thus, these interdependencies could lead to a correlation of dye spread with other gas properties without causation, which makes it difficult to pinpoint the cause of gas mixing in the CGM.

\begin{table}
\centering
\caption{This table shows the Spearman rank coefficient ($\rs$, to quantify correlation strength) between some averaged gas properties of the selected regions ($10 \times 10 \times 10 \kpc$ boxes). \rev{Results show significant correlations ($\rs \ge 0.6$) between $\vdis$ \rev{and} $\rho_{\rm over}$, $\KEdens$, $|\mathcal{S}_{ij}|$, and $|\Omega_{ij}|$} .}
\begin{tabular}{|c|c|c|}

\hline
Properties & $\rs$ \\
\hline
$\vdis$ vs $T$ & $0.28$ \\
$\vdis$ vs $Z/Z_{\odot}$ & $0.42$ \\
$\vdis$ vs $\rho_{\rm over}$ & $0.69$ \\
$\vdis$ vs $\KEdens$ & $0.63$ \\
$\vdis$ vs $B$ & $0.48$ \\
$\vdis$ vs $|\mathcal{S}_{ij}^*|$ & 0.91\\
$\vdis$ vs $|\Omega_{ij}|$ & 0.91\\
\hline

\end{tabular}
\label{tab:init_corr_table}
\end{table}

It can be seen from \autoref{fig:spread_gasprops} that out of all selected gas properties, velocity dispersion \rev{and symmetric shear tensor magnitudes have} the highest Spearman coefficient $\rs$ when correlating with dye spread. This suggests that \rev{velocities are} a critical factor in determining the rate of gas mixing. The strong dependence of velocity dispersion on other gas properties, as indicated by the high correlation coefficients in \autoref{tab:init_corr_table}, could explain the high $\rs$ values obtained for $\KEdens$ and $\rho$. Since more ordered flows ($i$, $o$, $s$) have a lower velocity dispersion, they mix less (or more slowly) with their surroundings. This is consistent with our findings in \autoref{fig:spread_colour}. \rev{Naturally, the tensor quantities created from partial derivatives of the velocity fields correlate with velocity dispersion as well. In fact, SPH codes often use $\vdis$ or $|\mathcal{S}_{ij}^*|$ for subgrid-scale models of turbulent mixing \citep{Wadsley2008, Williamson2016}. The latter comes from the Smagorinsky equation, which can be expressed as}

\begin{ceqn}
\begin{align}\label{eq:smagorinsky}
    D = (\mathrm{C_s}h)^2|\mathcal{S}_{ij}^*|,
\end{align}
\end{ceqn}

\rev{where $D$ is the Smagorinsky diffusivity, $\mathrm{C_s}$ is the Smagorinsky model constant, and $h$ is the resolution scale.}

\begin{figure*}
    \includegraphics[width=2\columnwidth]{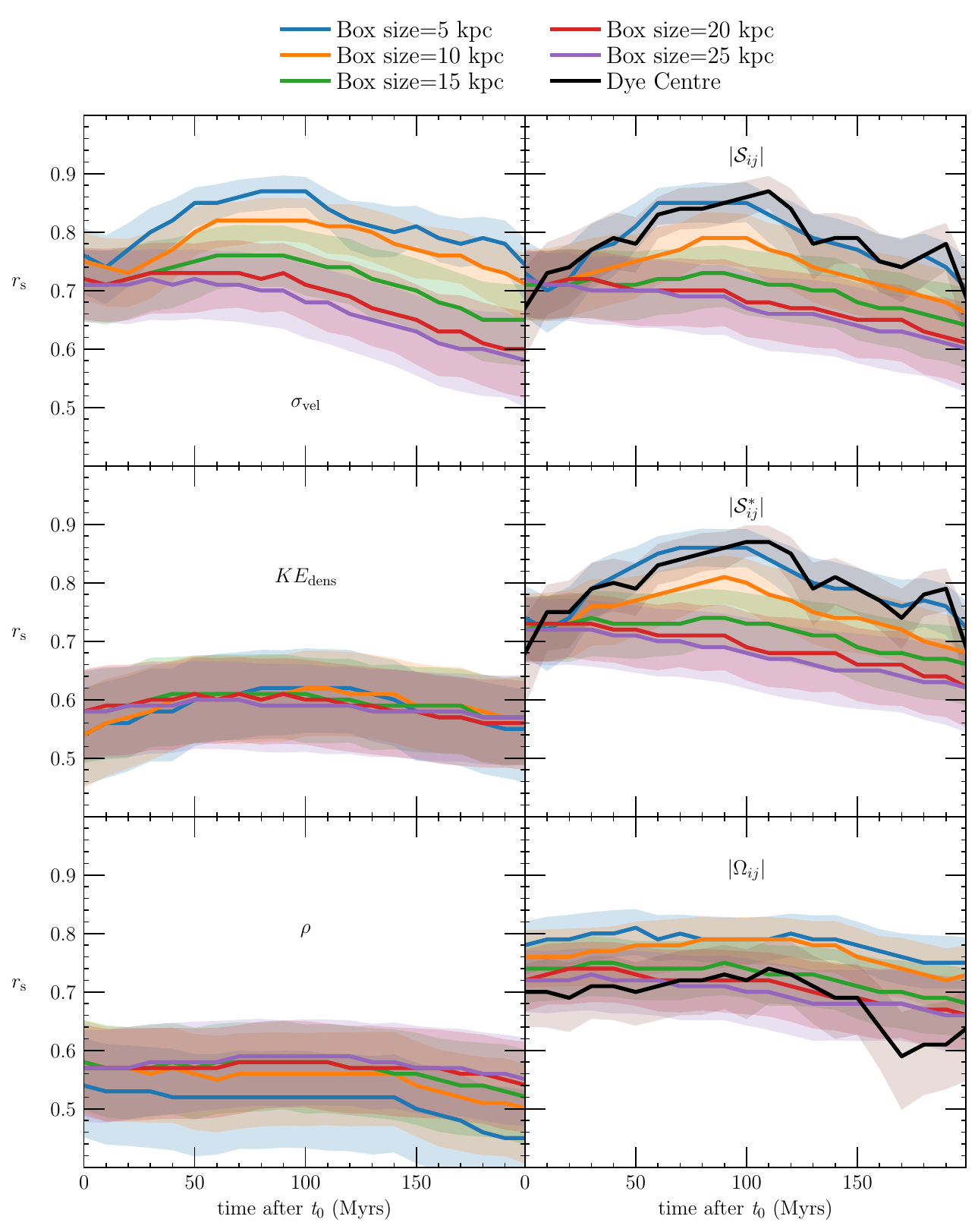} 
    \caption{Spearman rank coefficient ($\rs$) of the correlation between the magnitude of dye spread after $200 \Myr$ and averaged gas properties in differently sized boxes (see legend) around the injected dye at each simulation output time. The error bars represent $1\sigma$ confidence intervals from bootstrapping the sample $10^{4}$ times. Among the \rev{six} gas properties shown, $\vdis$\rev{, $|\mathcal{S}_{ij}|$, and $|\mathcal{S}_{ij}^*|$ provide} the highest $\rs \sim 0.86$\rev{. The maximum correlation for $\vdis$ and the tensor quantities occurs in a $5 \kpc$ box, and at the dye centre, respectively, at $\sim 100 \Myr$.} Other properties produce lower $\rs$ values and marginal changes with time and box size, suggesting an indirect correlation. }
    \label{fig:spread_time_order}
\end{figure*}

To fully understand the relationship between properties with higher $\rs$ and gas mixing and to identify other factors that may influence this process, we explore the spatial and temporal relationships of $\vdis$, $\KEdens$, $\rho$\rev{, and tensor quantities} at different times after dye injection with $|\spread|$ $200 \Myr$ after dye injection in \autoref{fig:spread_time_order}. Five different analysis box sizes ($5, 10, 15, 20, 25 \kpc$) are selected at all output times (from $\tnaught$, dye injection time, to $\tnaught+200 \Myr$, in $10 \Myr$ spacing) to a create a region around the 95 dye injected cells (where the box centre follows the cell with highest dye mass as the simulation evolves). Gas properties averaged in each analysis region are correlated with $|\spread|$ at $\tnaught+200 \Myr$ at every output time. \rev{The tensor quantity plots (right panels) contain the calculation performed at the dye centre as well, since the tensor quantities can be accurately computed there without the need for averaging.} The $\rs$ values are plotted in \autoref{fig:spread_time_order}. The error bars represent $\rm 1\sigma$ confidence levels from bootstrapping the $\rs$ samples $10^4$ times. 

The $\rs$ values change significantly with time and box size for $\vdis$ \rev{and symmetric tensor quantities} while remaining nearly constant for $\KEdens$ and $\rho$. The changes in the values of $\vdis$ are marginal with time and scale up with increasing box sizes (see \autoref{fig:init_prop_time} in \autoref{sec:appendix_misc}) because larger boxes \rev{usually} probe more of the random motions in the CGM. \rev{The presence of laminar shearing flows could also increase the velocity dispersion; however, the systematic increase of $\vdis$ with the box sizes suggests that laminar shearing flows are possibly in all directions, becoming more equivalent to turbulent motions.} Since the magnitudes of quantities do not affect Spearman rank correlation test results, scaled-up $\vdis$ values for different box sizes also should not affect $\rs$. However, the correlation of $\spread$ at $\tnaught+200 \Myr$ with these $\vdis$ samples changes significantly. On the other hand, the evolution of $\KEdens$ and $\rhoover$ is significant (the standard deviation increases by a factor of 2 over $200 \Myr$, see \autoref{fig:init_prop_time}). Despite that, their correlation with $\spread$ does not vary significantly (within $1\sigma$ error). These results suggest that the correlation between dye spread and $\vdis$ is stronger and more sensitive to variations than the correlation between dye spread and \rev{$\KEdens$ and $\rho$}. \rev{When applying the same arguments to the tensor quantities, we find that the dye spread correlates best with the symmetric parts. The traceless symmetric part exhibits very similar behaviour to just the symmetric part, suggesting that pure shear, rather than volumetric expansion/contraction, drives gas mixing, as expected. The correlation with the antisymmetric tensor is \rev{also strong, possibly due to its correlation with the $\vdis$ (see \autoref{tab:init_corr_table}).} Thus,} the high $\rs$ values (maximum $\sim 0.87$) indicate a strong dependence of $\spread$, and hence of gas mixing, on the velocity dispersion \rev{and the shear} of the surrounding medium.

As we increase the analysis box size, the overall $\rs$ at all times reduces significantly in the case of $\vdis$ \rev{and the tensor quantities} (see \rev{top-left and right panels} of \autoref{fig:spread_time_order}), suggesting a stronger dependence of $\spread$ (average $\spread$ for all dyes after $200 \Myr$ is $\approx 11 \kpc$) on smaller \rev{regions} surrounding the dyed areas. \rev{Velocity-derived quantities are the primary cause of the dye mixing.} Since larger boxes are more likely to have regions without dye, their velocity dispersion \rev{and average tensor quantities (tensor magnitudes averaged over the region) }would include undyed regions, causing the correlation \rev{with} $\spread$ to worsen. Additionally, the smaller boxes contain highly concentrated dye (see \autoref{fig:flashyfig}), so their \rev{velocity-derived quantities} also cause a larger change in dye spread compared to more diluted areas of the dye. Thus, \rev{the extent of the mixed gas (|$\spread$|) is best predicted with $\vdis$ and the symmetric shear tensors (see \autoref{eq:smagorinsky}) in smaller regions around the dye.}

\rev{We can draw more inferences from the time evolution of $\rs$ with $\vdis$ and symmetric shear tensors.} For the smaller \rev{regions}, a maximum \rev{always occurs} between $\approx \tnaught+70 \Myr$ and $\tnaught+120 \Myr$, implying that the mixing of gas from a cell with its surroundings after $200 \Myr$ is best defined by the \rev{velocity-derived quantities} between $\tnaught+ (70 \-- 120) \Myr$. This suggests a delay in the transfer of the local velocity dispersion \rev{and shear} to the spread of the dye. Because, if it were instantaneous, a continuous rise in $\rs$ would be expected. The initial increase in $\rs$ could be due to the dye gradually filling the box where \rev{the velocity-derived quantities are} calculated, such that \rev{their} transfer to the dye spread becomes more optimal. When we perform the same simulations with dye injection in $2 \times 2 \times 2 \kpc$ clumps (a collection of 8 cells around the original dye injection cells) instead of $1 \kpc$ cells, the results remain qualitatively the same (see \autoref{fig:clump_corr} in \autoref{sec:appendix_misc}). However, the location of the maximum shifts slightly towards earlier times, between $\tnaught=50 \textrm{ and } 100 \Myr$, perhaps due to the dye injected in clumps covering the box more quickly, thus reaching an optimal configuration for the transfer of $\vdis$ and \rev{shear} to dye spread sooner.

\subsection{Diffusion Coefficient}
\label{sec:diffusion_coefficient}
\rev{With our tracer dyes, we can calculate the average diffusion coefficient over $\rm T \ (\Myr \ )$ as}
\begin{ceqn}
\begin{align}\label{eq:DC}
    {\rm \langle DC \rangle_{T}} = |\spread|^2 \ / \ {\rm T \  (\Myr \ )},
\end{align}
\end{ceqn}

\begin{figure}
    \includegraphics[width=1\columnwidth]{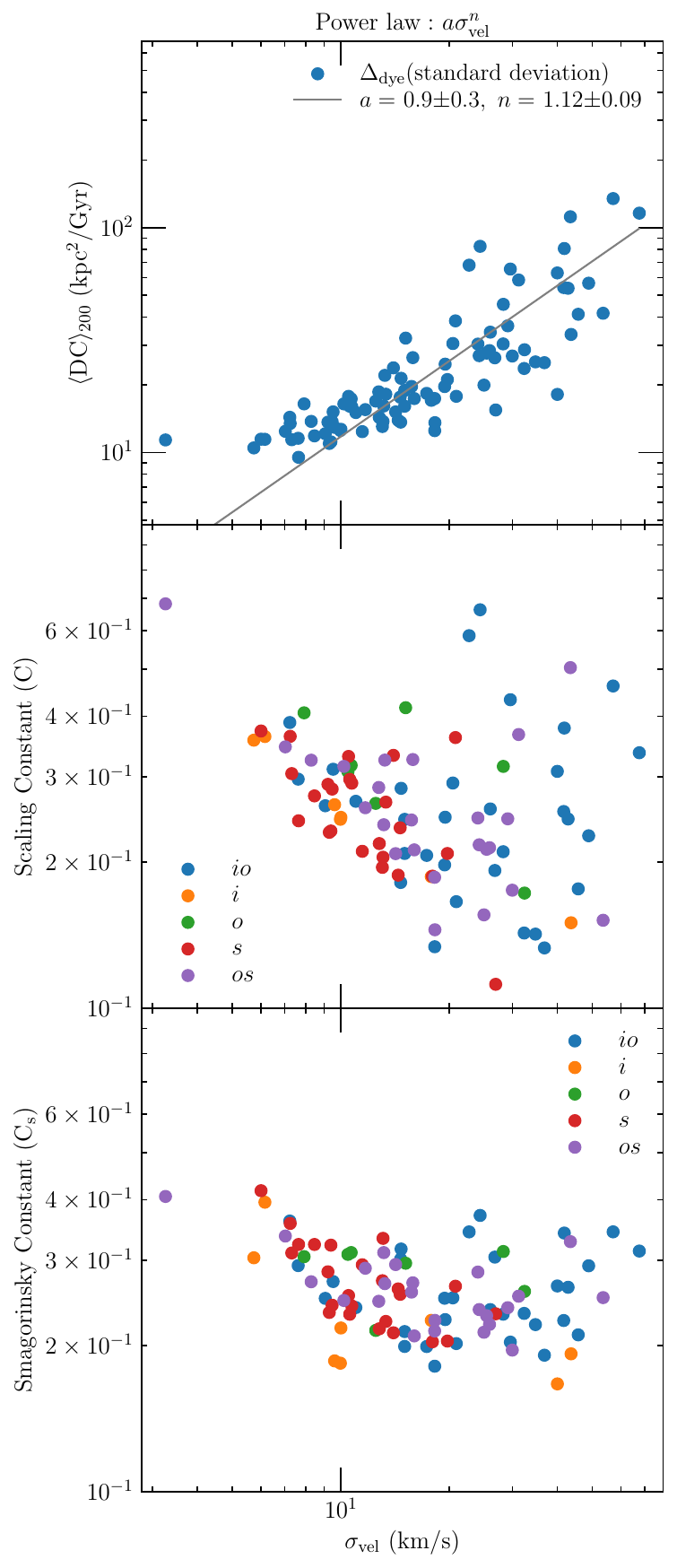} 
    \caption{The top panel shows the average diffusion coefficients for dyes over $200 \Myr$, calculated as $|\spread|^2/(200 \Myr)$, as a function of velocity dispersion. \rev{The standard deviation definition of} $\spread$\rev{ is used.} Black lines show a power law fit (parameters mentioned in the title and legend). The diffusion coefficients scale almost linearly ($n=1.1 \pm 0.09$) with the velocity dispersion. The \rev{middle and }bottom \rev{panels show} the scaling constant (C, computed using \autoref{eq:C}) \rev{and Smagorinsky constant (${\rm C_s}$, computed from \autoref{eq:Cs}), respectively. We derive these constants for our simulations using well-known subgrid diffusion coefficient estimation methods of SPH simulations. We find a good agreement between typical values in the literature and our values (both are approximately ${\rm C}=0.1-1$ and ${\rm C_s}=0.2-0.4$)}.}
    \label{fig:diffusion_coefficient}
\end{figure}

\rev{where we substitute $\rm T = 200 \Myr$, the timescale over which we track the dye. We use the standard deviation of dye spread $\spread$ throughout this section, as it aligns with the physics of diffusion and focuses on the bulk of the dye distribution rather than the tails. The resulting diffusion coefficients are shown in the top panel of \autoref{fig:diffusion_coefficient}. Assuming a power-law dependence of the diffusion coefficient on the velocity dispersion, we determine the power-law exponent to be $n \approx 1.1$ (almost linear)\footnote{\rev{We find similar power-law exponents, $n\sim 1.1$, for simulations with standard + 2 kpc refinement.}}.}

\rev{We then correlate this diffusion coefficient with} the velocity dispersion \rev{and symmetric shear tensors (based on results from \autoref{sec:mag_gas_mixing}). SPH codes also use similar models for diffusivity in their subgrid models \citep{Klessen2003, Wadsley2008, Greif2009, Shen2013, Williamson2016, Colbrook2017, Rennehan2019, Rennehan2021}}. \rev{Some studies estimate that the diffusion coefficient} $={\rm C}\,\vdis\,l_{\rm turb}$\rev{, where} $\rm C$ is the scaling constant \citep[usually assumed to vary from $0.1 \-- 1$, as seen in][]{Wadsley2008}, and $l_{\rm turb}$ is turbulent scale relevant for gas mixing (they assume it to be equal to the simulation resolution). \rev{Given that the dye spread allows us to calculate the actual diffusion coefficient in our simulations (\autoref{eq:DC}), we can estimate the scaling constant ${\rm C}$ from the SPH definition for our simulation as}

\begin{ceqn}
\begin{align}\label{eq:C}
    {\rm C} = |\spread|^2/(200 \Myr\,\vdis\,l_{\rm turb}), 
\end{align}
\end{ceqn}

where $\rm C$ (scaling constant) is a factor that scales \rev{the} estimated diffusion coefficient ($\vdis\,l_{\rm turb}$) to the actual diffusion coefficient (\autoref{eq:DC})\rev{,} $l_{\rm turb}$ is the approximate average length scale of dye evolution over $200 \Myr$ ($5 \kpc$), and $\vdis$ is the velocity dispersion around our dyes at $l_{\rm turb}$ scales. \rev{Because} $|\spread|$ after $200 \Myr$ correlates the best with velocity dispersion in regions of $5 \kpc$ (see \autoref{fig:spread_time_order})\rev{,} it is a reasonable \rev{to assume} $l_{\rm turb}=5 \kpc$. It should be noted that turbulence can often be driven on much larger scales (a few tens to hundreds of kpc) in the CGM \citep[see Fig. 8 in][]{Pakmor2020}. However, the turbulent length scales responsible for dye mixing are smaller than or similar to the extent of dye spread, cascaded down from large-scale turbulence.

\rev{Similar to the scaling constant derivation above, we can determine the Smagorinsky constant $\rm C_s$ by estimating the diffusivity from the Smagorinsky equation (\autoref{eq:smagorinsky}). In this case,}

\begin{ceqn}
\begin{align}\label{eq:Cs}
    {\rm C_s}^2 = |\spread|^2/(200 \Myr\,|\mathcal{S}_{ij}^*|\,l_{\rm turb}^2), 
\end{align}
\end{ceqn}

\rev{where we replace the resolution scale $h$ with the approximate turbulent length scale over $200 \Myr$ ($l_{\rm turb} = 5 \kpc$). The middle and bottom panels of \autoref{fig:diffusion_coefficient} show the scaling constant ${\rm C}$ and Smagorinsky constant ${\rm C_s}$ for our simulation, respectively, for different flow types. $\rm C$ ranges $0.09-0.7$, similar to scaling constants used in subgrid models in SPH simulations, ranging from $0.1 - 1$ \citep{Williamson2016}. ${\rm C_s}$ in our simulation ranges from 0.2--0.4 (see bottom panel of \autoref{fig:diffusion_coefficient}). These values are similar to those found in the literature -- ${\rm C_s = 0.2}$ \citep{Clark1979}, ${\rm C_s = 0.1-0.2}$ \citep{Garnier2009}, ${\rm C_s = 0.37}$ \citep{Shen2010, Shen2013, Brook2014}, and ${\rm C_s = 0.29}$ \citep{Wadsley2017}}\footnote{\rev{Note that there are many different definitions of the Smagorinsky constant in the literature; \citet{Rennehan2019} performed the correct conversions of other definitions to the ${\rm C_s}$ we use.}}\rev{. Moreover, we do not see any dependency of these constants on different flow types.}

\rev{Obtaining constants ${\rm C}$ and ${\rm C_s}$ similar to the values in the literature is promising for both the implicit model of mixing in our simulation and the typical subgrid models used in certain SPH simulations. However, the implications require careful consideration.} The diffusion coefficient of the dye is a time-varying quantity for our simulations. As the dye evolves, the relevant $\vdis$ and $l_{\rm turb}$ for estimating the diffusion coefficient also change. Since we do not account for this \rev{in this section}, reported scaling constants are approximately the average scaling constants for gas mixing over $200 \Myr$ (which are for a time step in subgrid mixing models). \rev{In addition,} numerical mixing is present in our simulations (see \autoref{Sec:num_mix}). Thus, more detailed time-dependent analysis at different resolutions is required to disentangle these factors, which is beyond the scope of this study.

\subsection{Direction of Gas Mixing}
\label{sec:dir_gas_mixing}
\begin{figure}
    \includegraphics[width=1\columnwidth]{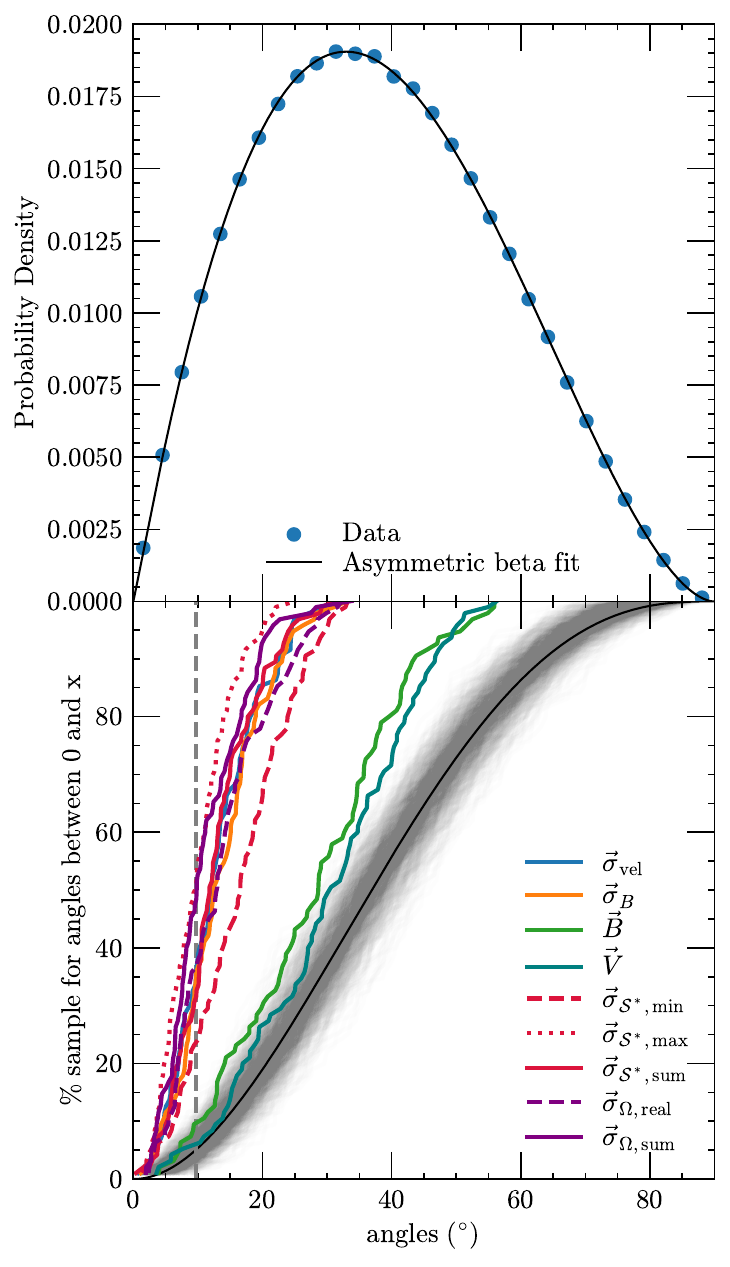} 
    \caption{The top panel shows the PDF of possible angles between two random vectors in the positive 3D octant. The black curve shows an \rev{asymmetric beta function} fit (for smoothing) to blue data points (generated using $10^5$ uniformly-spaced random vector pairs). The \rev{bottom panel shows} the cumulative distribution functions (CDF) as percentages corresponding to alignments within x-axis angles. The black curves \rev{are} the control sample CDFs, created from the top panel fit to the PDF. The coloured curves \rev{in the bottom panel} correspond to angles between $\vec{\spread}$ and gas properties \rev{$\vec{\vdis}$, $\vec{\sigma}_{B}$, $\vec{B}$, $\vec{V}$, $\vec{\sigma}_{\rm \mathcal{S}^*, min}$, $\vec{\sigma}_{\rm \mathcal{S}^*, max}$, $\vec{\sigma}_{\rm \mathcal{S}^*, sum}$, and $\vec{\sigma}_{\rm \Omega, real}$}. 5 per cent of the random vectors lie within \rev{9.6} degrees of one another\rev{, as shown by the grey dashed line}. \rev{All the selected} gas properties exhibit a higher directional correlation with $\vec{\spread}$ than the random control sample.}
    \label{fig:oct1_angles}
\end{figure}

\begin{figure*}
    \includegraphics[width=2\columnwidth]{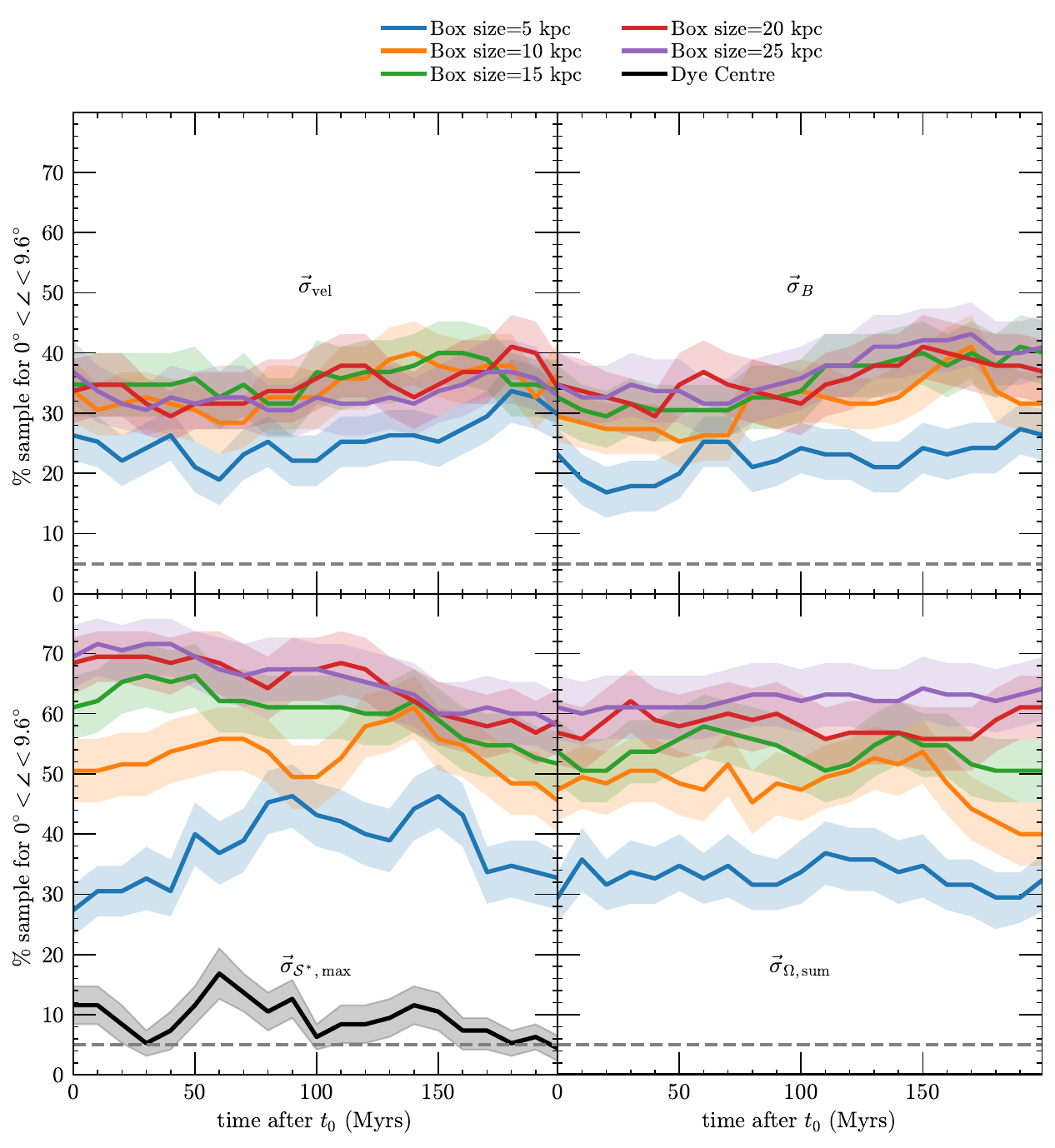} 
            \caption{The percentage \rev{of dyes, $\vec{\spread}$, with the direction of dye spread (after $200 \Myr$) aligned within $9.6^{\circ}$ of} \rev{$\vec{\vdis}$ (top-left panel), $\vec{\sigma}_{B}$ (top-right panel), $\vec{\sigma}_{\rm \mathcal{S}^*, max}$ (bottom-left panel) and $\vec{\sigma}_{\rm \Omega, sum}$ (bottom-right panel), computed} in different analysis box sizes as per the legend, as a function of time. The grey dashed line (5\%) represents the control sample \rev{(same as the one in \autoref{fig:oct1_angles})}. Larger boxes tend to show stronger alignment than the smallest box ($5 \kpc$).\rev{ Quantities associated with the dispersion of stretching and plane of rotation eigenvectors exhibit a higher alignment with the dye shape than the dispersion of velocities and magnetic fields.}}
    \label{fig:spread_dir_time_order}
\end{figure*}

In the previous subsection, we discussed the dependence of the magnitude of gas mixing on physical properties. Here, we analyse the directional dependence of gas mixing on physical properties. To do this, we correlate the spread vector ($\vec{\spread} = \spreadx \hat{i} + \spready \hat{j} + \spreadz \hat{k}$) with several physical property vectors -- velocity dispersion ($\vec{\vdis} = \sigma_{V_{\rm x}} \hat{i} + \sigma_{V_{\rm y}} \hat{j} + \sigma_{V_{\rm z}} \hat{k}$), \rev{$\vec{V}, \vec{B}$, and magnetic field dispersion computed from standard deviations of magnetic field vector components ($\vec{\sigma}_B = \sigma_{B_{\rm x}} \hat{i} + \sigma_{B_{\rm y}} \hat{j} + \sigma_{B_{\rm z}} \hat{k}$)}. \rev{To compare with the tensor quantities, we use combinations of tensor eigenvalues and eigenvectors to create vectors with physical meanings. Since quantities derived from symmetric and traceless symmetric tensors yield similar results, we only show results from the latter. For the traceless symmetric ($\mathcal{S}_{ij}^*$) shear tensor, we construct three vectors each as follows.}

\begin{ceqn}
    \begin{align}\label{eq:symmangles}
        \vecsymmtrmin = \lambda_{\rm min}\vec{E}_{\rm min}, \\
        \vecsymmtrmax = \lambda_{\rm max}\vec{E}_{\rm max}, \\
        \vecsymmtrsum = \Sigma_{n} \lambda_{n}\vec{E}_{n}. 
    \end{align}
\end{ceqn}

\rev{Here, $\lambda$ and $\vec{E}$ represent $n$ eigenvalues and the corresponding unit eigenvectors of symmetric tensors, respectively. $\lambda_{\rm min/max}$ and $\vec{E}_{\rm min/max}$ are the minimum/maximum eigenvalues and the corresponding eigenvectors, respectively. The symmetric shear tensors always have three real eigenvalues with at least one positive and one negative value. Importantly, the positive and negative eigenvalues represent the strengths of deformation, either stretching or compression, without any volume change, respectively. The corresponding eigenvectors point in the directions of stretching and compression. Note that deformation (changes in shape) is distinct from volumetric changes (expansion or contraction) in the fluid. Since both symmetric and traceless symmetric shear tensors yield similar results, we argue that volumetric changes, compared to deformation, play a minimal role in mixing the dyes in our simulations. Similarly, for antisymmetric shear tensors, we create two vectors as follows.}

\begin{ceqn}
    \begin{align}
        \vecantisymmreal = \lambda_{\rm real}\vec{E}_{\rm real}, \\
        \vecantisymmsum = \Sigma_{n} \lambda_{n}\vec{E}_{n}. \label{eq:antisymmangles}
    \end{align}
\end{ceqn}

\rev{Since the antisymmetric shear tensor has one real eigenvalue $\lambda_{\rm real}$ ($\sim 0$), we use it with the corresponding eigenvector $\vec{E}_{\rm real}$ to create $\vecantisymmreal$, which points towards the rotational axis. The imaginary eigenvalues of a real antisymmetric matrix are complex conjugates, and their corresponding eigenvectors are also complex conjugates (representing the plane of rotation); therefore, the vector created by the sum of all the eigenvalues and eigenvectors $\vecantisymmsum$ is also real. Since $\lambda_{\rm real} \sim 0$, the summation leads to $\vecantisymmsum$ pointing in the direction perpendicular to the rotational axis, i.e., the plane of rotation. Similar to the calculation of the velocity and magnetic field dispersion in boxes, we take the standard deviation of the three components of the vectors constructed above from tensors, and show the resultant vectors as $\vec{\sigma}_{\rm \mathcal{S}^*, min}$, $\vec{\sigma}_{\rm \mathcal{S}^*, max}$, $\vec{\sigma}_{\rm \mathcal{S}^*, sum}$, $\vec{\sigma}_{\rm \Omega, real}$, and $\vec{\sigma}_{\rm \Omega, sum}$. }

\rev{We now have vectors for all the quantities of interest. Typically, vectors can point in any direction; however, all the vectors we construct only have positive components\footnote{\rev{If the vectors do not have positive components, as in the case of $\vec{B}$ and $\vec{V}$, we take the absolute values of their components for this analysis.}}, thus, they point in a single positive octant. Then we compute the angle between $\vec{\spread}$ and these positive component vectors.} The distribution of angles between random positive vectors is different from full 3D vectors. Thus, we generate a control sample by calculating angles between two random vectors\rev{, distributed uniformly in spherical geometry,} free to point anywhere in one octant (with all positive values). The top panel of \autoref{fig:oct1_angles} shows the probability density of the angles between pairs of these random vectors ($\theta = [0, \pi/2]$). \rev{The data is fitted to an asymmetric beta function}

\begin{ceqn}
\begin{align}
    f(x; \alpha, \beta_b, k) = \frac{x^{\alpha} (90 - x)^{\beta_b} \exp(-k x)}{90^{\alpha + \beta_b + 1} \; B(\alpha + 1, \beta_b + 1)},
\end{align}
\end{ceqn}

\rev{where, }

\begin{ceqn}
\begin{align}
    B(a, b) = \int_{0}^{1} t^{a-1} (1-t)^{b-1} \, dt,
\end{align}
\end{ceqn}

\rev{is the beta function. The best-fit parameters are $\alpha=1.04, \beta_b=1.80$, and $k=1.7 \times 10^{-5}$. The fitting is done to remove} the statistical variations due to a finite number of data points. This control sample represents the distribution in a set of random uncorrelated positive vectors and can be compared to the distribution of angles between $\vec{\spread}$ and vectors representing physical properties of the gas.


The \rev{bottom} panel of \autoref{fig:oct1_angles} shows the cumulative distribution functions (CDF) (as percentages) of the above control sample PDF (solid black curve) and the angles between $\vec{\spread}$ and \rev{$\vec{\vdis}$, $\vec{\sigma}_{B}$, $\vec{B}$, $\vec{V}$, $\vec{\sigma}_{\rm \mathcal{S}^*, min}$, $\vec{\sigma}_{\rm \mathcal{S}^*, max}$, $\vec{\sigma}_{\rm \mathcal{S}^*, sum}$, and $\vec{\sigma}_{\rm \Omega, real}$}. The grey-shaded region \rev{in the bottom panel represents} $1000$ different CDFs with 95 random data points (equal to the number of dyes) from the control sample to showcase statistical variations due to a limited number of available data points. \rev{All} vectors are computed at the dye injection time in a $10 \kpc$ box around the injection location \rev{$100 \Myr$ after injection}, and $\vec{\spread}$ is calculated $200 \Myr$ after dye injection. The physical property CDFs are created from the 95 dye injection locations. 

Compared to the control sample, \rev{all properties have more alignment with $\vec{\spread}$ as they have higher CDFs}. \rev{In} \autoref{tab:angle_corr_table}, \rev{we quantify} the average angle between dye spread and physical properties. \rev{All the average angles are smaller than the control sample average angle, with $\vec{\sigma}_{\rm \mathcal{S}^*, max}$ exhibiting the highest alignment. This suggests that the dyed cloud's shape is determined by the dispersion of stretching eigenvectors. Put simply, the direction with the greatest variance in fluid stretching deformation is where the most gas mixing occurs. The high angular alignment of the dye spread with $\vec{\sigma}_B$ and $\vec{\sigma}_{\rm vel}$ suggests that magnetic fields and velocity fields are also aligned within themselves \citep{Seta2020, Seta2021}. When we performed the same experiment with the positive components of the averages of these vectors rather than the standard deviation (see $\vec{B}$ and $\vec{V}$ in \autoref{fig:oct1_angles}), we found that the level of alignment is closer to the control sample, suggesting that dispersion is a better measure of the diffusive processes. It is difficult to pinpoint the best aligning physical quantity by analysing just one region size and time, thus, we perform the spatial and temporal analysis in the same way as \autoref{fig:spread_time_order} for some of the best aligning vectors from \autoref{fig:oct1_angles}.}

\begin{table}
\centering
\caption{\rev{This table shows the mean angle and its standard deviation between dye spread and the physical properties $\vec{\vdis}$, $\vec{\sigma}_{B}$, $\vec{V}$, $\vec{B}$, $\vec{\sigma}_{\rm \mathcal{S}^*, min}$, $\vec{\sigma}_{\rm \mathcal{S}^*, max}$, $\vec{\sigma}_{\rm \mathcal{S}^*, sum}$, $\vec{\sigma}_{\rm \Omega, real}$, $\vec{\sigma}_{\rm \Omega, sum}$, as well as the control sample.}}
\begin{tabular}{|c|c|c|}

\hline
Properties & Angle ($^\circ$) \\
\hline
$\vec{\vdis}$ & $12.96 \pm 3.35$ \\
$\vec{\sigma}_{B}$ & $13.62 \pm 3.42$ \\
$\vec{V}$ & $30.43 \pm 6.39$ \\
$\vec{B}$ & $28.19 \pm 6.49$ \\
$\vec{\sigma}_{\rm \mathcal{S}^*, min}$ & $16.56 \pm 3.9$ \\
$\vec{\sigma}_{\rm \mathcal{S}^*, max}$ & $9.96 \pm 2.63$ \\
$\vec{\sigma}_{\rm \mathcal{S}^*, sum}$ & $12.77 \pm 3.21$ \\
$\vec{\sigma}_{\rm \Omega, real}$ & $13.69 \pm 3.68$ \\
$\vec{\sigma}_{\rm \Omega, sum}$ & $11.22 \pm 3.14$ \\
Control Sample & $37.90 \pm 9.20$\\
\hline

\end{tabular}
\label{tab:angle_corr_table}
\end{table}

\autoref{fig:spread_dir_time_order} shows the percentage of the sample for angles lower than \rev{$9.6^{\circ}$} (corresponding to 5 per cent of the control sample\rev{, shown as grey dashed lines in \autoref{fig:oct1_angles} and \autoref{fig:spread_dir_time_order}}) \rev{between $\vec{\spread}$ and $\vec{\vdis}$ (top-left panel), $\vec{\sigma}_{B}$ (top-right panel), $\vec{\sigma}_{\rm \mathcal{S}^*, max}$ (bottom-left panel) and $\vec{\sigma}_{\rm \Omega, sum}$ (bottom-right panel)}. This is computed for $\vec{\spread}$ after $200 \Myr$ for physical properties inside differently sized analysis boxes at every simulation output time. \rev{Every quantity shows a clear }alignment with dye spread, regardless of box size or time \rev{and this} result does not qualitatively depend on our choice of $9.6^{\circ}$. \rev{Generally, smaller boxes show} a weaker correlation than larger boxes. \rev{Moreover, a higher per cent of the dyes ($\sim 60\%$) are more aligned (angles $\leq 9.6^{\circ}$) with $\vec{\sigma}_{\rm \mathcal{S}^*, max}$ and $\vec{\sigma}_{\rm \Omega, sum}$ compared to $\vec{\vdis}$ and $\vec{\sigma}_{B}$ ($\sim 35\%$)}. \rev{Thus, the dye shape is best determined by the dispersion of stretching (deformation) and plane of rotation eigenvectors} in boxes spanning the dye volume (box size $\ge 10 \kpc$) unlike dye spread magnitudes which correlated the best with boxes containing high dye mass concentrations (box size $\sim 5 \kpc$). The average dye spread is $\sim 11 \kpc$ after $200 \Myr$. Larger boxes ($\ge 10 \kpc$) correlating better with the dye spread shape suggests that the correlation forms later in the evolution. Since the best correlation for dye spread magnitude is between $70-120 \Myr$ after injection \rev{(see \autoref{fig:spread_time_order})}, the spread shape probably changes faster than the spread magnitude. However, \rev{the temporal trend for alignment analysis is statistically insignificant}, because variations \rev{with time} in \rev{\autoref{fig:spread_dir_time_order}} are within error bars. \rev{Combining the alignment results with our results from \autoref{sec:mag_gas_mixing} and \autoref{fig:spread_time_order}, we conclude that the traceless symmetric shear tensor best predicts both the magnitude and shape of gas mixing in our simulations.}

\subsection{Radial profile of gas mixing}

\begin{figure}
    \includegraphics[width=1\columnwidth]{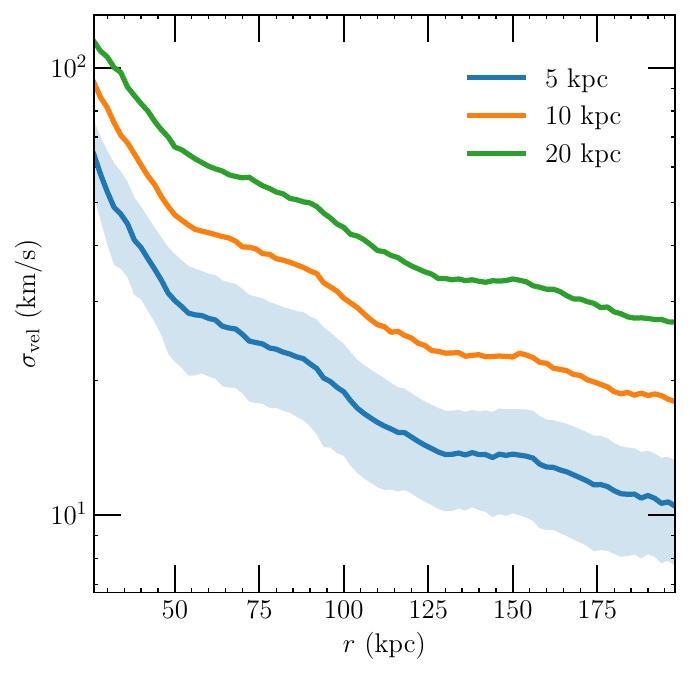}
    \caption{Velocity dispersion as a function of distance from the galactic centre ($r$). Three different characteristic length scales/diameters of the spherical regions ($5, 10, 20 \kpc$) are used to calculate $\vdis$ by bootstrapping 100 samples of 100 data points at each $r$. The shaded blue region shows the scatter in $\vdis$ when evaluated on 5~kpc scales. $\vdis$ decreases radially and scales up with increasing characteristic lengths. }
    \label{fig:vdis_ld}
\end{figure}

As established in the earlier sections, the amount of gas mixing in the CGM of galaxies is heavily influenced by the velocity dispersion of the local surroundings. \autoref{fig:vdis_ld} shows the velocity dispersion as a function of distance from the galactic centre ($r$) for three spherical regions, with diameters 5, 10, and 20~kpc. To calculate velocity dispersion at a particular radius, all cells with a radial distance between $r-0.5 \kpc$ and $r+0.5 \kpc$ are selected. Out of these, 100 bootstrap samples containing 100 random spherical regions are generated. Spheres with a diameter equal to the region size are created around all these cells. The velocity dispersion is calculated by the volume-weighted standard deviation of the velocities of cells in these spheres. The median of 100 random cells is taken for each sample, and the median of 100 samples is reported as the velocity dispersion at that $r$. The shaded blue region represents the standard deviation of the velocity dispersion values for $5 \kpc$ regions. All cells were used to compute this scatter instead of bootstrapping, making it computationally prohibitive to perform the same calculation for the $10$ and $20 \kpc$ regions. The standard deviation for the $10$ and $20 \kpc$ regions is expected to scale up the same as their velocity dispersion values. 

The velocity dispersion decreases with radius, possibly due to higher gas flow activity happening closer to the centre because of outflows driven by stellar or AGN feedback, gas stripping from satellite galaxies, and higher inflow velocities \citep{vdv2021}. $\vdis$ is also larger for larger regions because large regions sample a more variable environment, as expected for turbulent flows \citep{Larson1981, Heyer2004, Federrath2012}. From \autoref{fig:vdis_ld}, we find the relation $\vdis \propto l_{\rm d}^{0.4-0.6}$, where $l_{\rm d}$ is the scale length of the region, i.e., the diameter of the region in which $\vdis$ is calculated. The scaling exponent is close to $1/2$, which is the value expected for supersonic turbulence \citep{KritsukEtAl2007,FederrathDuvalKlessenSchmidtMacLow2010,FederrathEtAl2021}.

The radial decrease of the velocity dispersion should cause a reduction in the amount of gas mixing in the CGM. Additionally, other properties of the CGM, such as metallicity and magnetic fields, also exhibit decreasing trends radially \citep[see Fig. 6 in][]{vdv2021}. Metals and magnetic fields originate primarily from the central galaxy and are ejected outwards in the galactic wind \citep{Christensen2018}. A lot of this material within a few tens of kiloparsecs from the galaxy centre is recycled back to the ISM \citep{Grand2019}. Additionally, the remaining material mixes more with regions closer to the galaxy centre and less with regions farther away due to the global velocity dispersion structure (\autoref{fig:vdis_ld}). A combination of these effects could create radially decreasing trends. Additionally, metal diffusion controls the distribution of metals on smaller scales, which depends on the turbulent gas mixing and thus velocity dispersion. Thus, understanding global velocity dispersion structure could be key in studying the chemical evolution of galaxies and the distribution of metals and magnetic fields in the CGM. 

\subsection{Numerical aspects and nature of gas mixing}
\label{Sec:num_mix}
To understand the nature of mixing, it is essential to determine the numerical effects of the necessarily limited resolution in these simulations. The spatial resolution of the order of 1 kpc, can limit the ability to capture the multi-phase nature of the CGM \citep{Gronnow2018, Ji2019, Das2023}, and thereby, small-scale physics (such as turbulence, magnetic fields, and fragmentation). Therefore, we study the convergence behaviour of the gas mixing in our simulations.

\begin{figure}
\begin{subfigure}{0.47\textwidth}
\centering
\includegraphics[width=1\columnwidth]{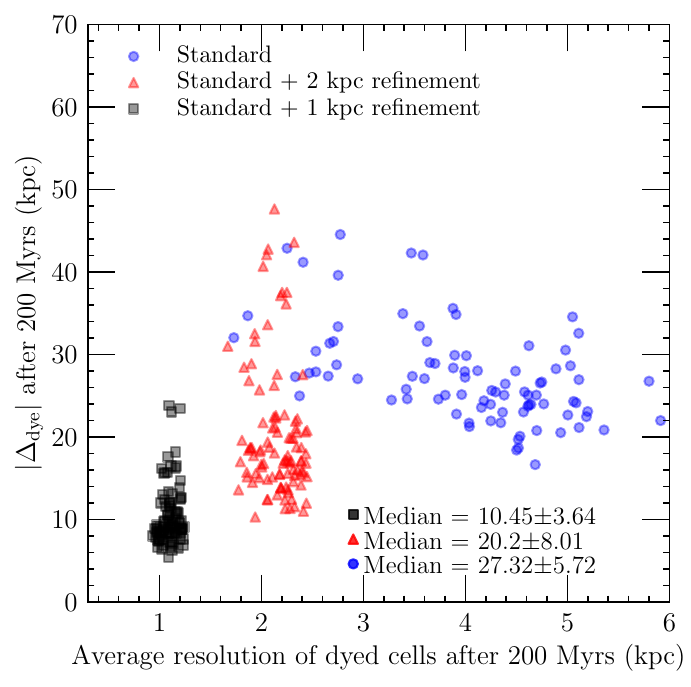} 
\end{subfigure}
\begin{subfigure}{0.47\textwidth}
\includegraphics[width=1\columnwidth]{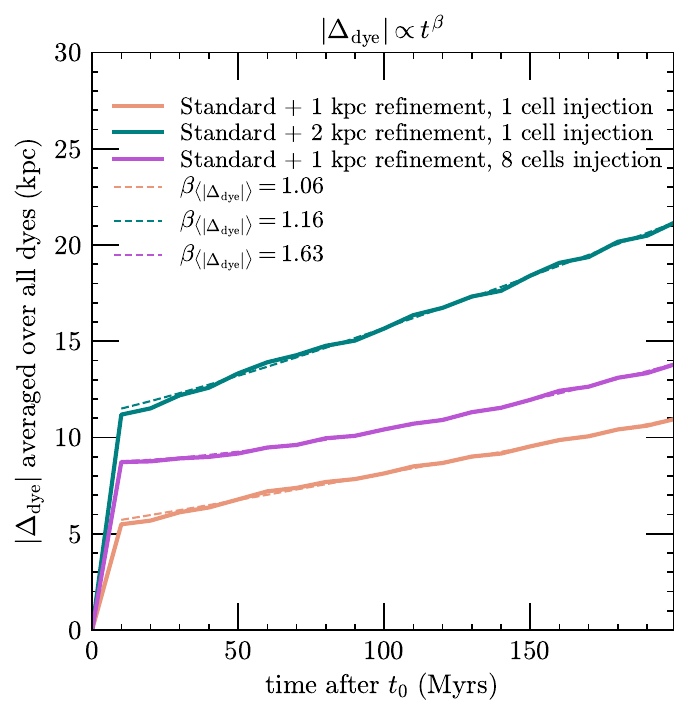} 
\end{subfigure}
\caption{The top panel shows the magnitude of spread for all dyes $200 \Myr$ after injection in different simulations as a function of the average resolution of dyed cells (dye mass fraction $>10^{-7}$). Three different resolution simulations -- standard (mass refinement), standard + 2 kpc spatial refinement, and standard + 1 kpc spatial refinement -- are compared to study the effect of resolution on gas mixing. The legend in the bottom right corner shows each simulation's median of all $|\spread|$ values. An increasing trend in $|\spread|$ is seen with decreasing resolution. \rev{Note that throughout the paper, we have used $\spread$ from the standard + 1 kpc run (black points). The $\spread$ for worse resolution cases is higher due to a higher velocity dispersion and larger impact of numerical mixing at those resolutions.} The bottom panel shows the evolution of $|\spread|$ averaged over all dyes. The same 1 kpc and 2 kpc simulation as in the top panel and an additional simulation with 1 kpc spatial refinement and a dye injection region of 8 neighbouring cells (as opposed to a single cell) are used to study the effects of both the size of the initial dye region and the resolution. A power law (mentioned in the title) is fitted to $|\spread|$ between $10-200 \Myr$ (to avoid the initial jump) by displacing $|\spread|$ such that it is zero at $10 \Myr$. The legend shows the exponent ($\beta_{\langle |\spread| \rangle}$) values for all three cases.}
\label{fig:num_mix}
\end{figure}

To do this, we resimulate tracer dyes in the same halo with standard + 2 kpcs spatial refinement and standard resolution (mass refinement only). The selection of dye locations follows the same methodology described in \autoref{sec:dye_location_selection} to ensure choosing diverse environments for injecting the dye. The top panel of \autoref{fig:num_mix} shows $|\spread|$ against the average resolution of dyed cells after $200 \Myr$ for all three simulations. The standard + 1 kpc and standard + 2 kpc runs have an almost fixed spatial resolution of 1 and 2 kpc, respectively, whereas the standard simulation has a range of cell sizes depending on their density.

A clear increase in $|\spread|$ is seen with worse resolutions in \autoref{fig:num_mix}. The median of $|\spread|$ values for each resolution (shown in the bottom right of \autoref{fig:num_mix}'s top panel) also show this increasing trend. When going from $1 \kpc$ resolution to $2 \kpc$, $|\spread|$ also increases twofold. This means that we are in a regime where the gas mixing is significantly affected by the numerical diffusion. That is why it is imperative to use the nearly fixed spatial resolution as we did here so that the dependence of gas mixing on physical properties is not due to spatial resolution variations that correlate with density in a mass refinement-only simulation. The standard resolution varies from $2-6 \kpc$ but $|\spread|$ shows a decreasing trend with lower spatial resolution (larger cell sizes). Because this simulation is refined on mass alone, the lower-resolution cells have lower densities. As we saw in \autoref{tab:init_corr_table}, densities are correlated with $\vdis$. Thus, lower resolution cells are on average located in regions with lower $\vdis$, which leads to less mixing and balances the increase in numerical mixing.

The bottom panel of \autoref{fig:num_mix} shows the evolution of $|\spread|$ averaged over all dyes in each simulation over $200 \Myr$. To check the dependence of dye spread on the initial injection, we repeat our standard + 1 kpc simulation but dye 8 neighbouring cells instead of a single cell as in our fiducial simulation, creating $(2 \kpc)^3$ clumps (similar to one cell at standard+2 kpc resolution) at $\sim 1 \kpc$ resolution. The jump in $|\spread|$ seen in \autoref{fig:num_mix} at the first output time is due to the transfer of a significant amount of dye from the injection cells to adjacent cells, perhaps due to steep gradients in dye concentrations leading to higher numerical diffusion. Afterwards, the average $|\spread|$, denoted as $\langle |\spread| \rangle$, is proportional to $t^{\beta_{\langle |\spread| \rangle}}$, where $\beta_{\langle |\spread| \rangle}=1.06, 1.16\textrm{, and } 1.63$ for one cell injection at $1 \kpc$, one cell injection at $2 \kpc$, and eight-cell injection at $1 \kpc$, respectively. We also look at the median and mode of $\beta$ values for the individual $95$ dyes in our sample. Mode is calculated as the most probable $\beta$ from the PDF of $\beta$ with a bin size of $0.36$. All the values are summarised in \autoref{tab:beta_table}.

\begin{table}
\centering
\caption{This table shows the statistics of $\beta$ for different simulations. (1 kpc, 1 cell), (2 kpc, 1 cell), and (1 kpc, 8 cells) refer to simulations with standard + 1 kpc refinement with 1 cell injection, standard + 2 kpc refinement with 1 cell injection, and standard + 1 kpc refinement with 8 cells injection, respectively. $\beta_{\langle |\spread| \rangle}$ is $\beta$ calculated from the average dye spread, $\beta_{\rm Median}$ is median of $\beta$ from all dye spreads, and $\beta_{\rm Mode}$ is the most probable $\beta$ from all dye spreads assuming a bin size of $0.36$ to create the PDF of $\beta$ values. }
\begin{tabular}{|c|c|c|c|}

\hline
Simulation & $\beta_{\langle |\spread| \rangle}$ & $\beta_{\rm Median}$ & $\beta_{\rm Mode}$ \\
\hline
(1 kpc, 1 cell) & $1.06$ & $0.94$ & $0.76 \pm 0.18$\\
(2 kpc, 1 cell) & $1.16$ & $1.03$ & $0.82 \pm 0.18$\\
(1 kpc, 8 cells) & $1.63$ & $1.42$ & $1.16 \pm 0.18$\\
\hline

\end{tabular}
\label{tab:beta_table}
\end{table}

For a normal diffusion (Brownian) process, $\beta = 0.5$. $\beta > 0.5$ suggests superdiffusion, and $\beta < 0.5$ suggests subdiffusion \citep{Bouchaud1990, Brandenburg2004, Bakunin2008}. Our $\beta$ values (\autoref{tab:beta_table}) ranging from $\sim 0.7-1.6$ suggest that our simulations are strongly superdiffusive and in extreme cases hyperballistic ($\beta > 1.0$) over $200 \Myr$ timescales \citep{Burnecki2015, Ilievski2021, Suzuki2022}. Thus, the diffusion coefficient is not a constant quantity with time as it is in the case of normal diffusion. Such behaviour is expected from turbulent environments and large-scale coherent flow structures \citep{Miesch2000}, both present in the CGM. Additionally, the presence of numerical diffusion and limited dye statistics could also affect $\beta$. We find that some dyes have $\beta > 4$ ($\sim 4$ per cent sample), leading to $\beta_{\langle |\spread| \rangle} > \beta_{\rm Median} > \beta_{\rm Mode}$. This is caused by dyes with faster spread at later times due to a change in the environment (an increase in the velocity dispersion) rather than the nature of mixing. Thus, inferring the nature of mixing from the most probable or median $\beta$ is more appropriate than using the average $\beta$ skewed by some outliers.  It should be noted that this diffusion behaviour could change when averaged over longer periods with more dyes. Thus, the superdiffusive behaviour of the dye might not apply to the general CGM.

\rev{\citet{Colbrook2017}, for supersonic ISM turbulence, found a similar time-independent scale-dependent diffusivity $\kappa \approx 0.5 \mathcal{M}_{\rm L_{box}}(k\mathrm{L_{box}})^{-\alpha_{c}}$, where wavenumber $k=2\pi/l$ describes variance on different scales, and $\mathrm{L_{box}}$ is the box scale at which the Mach number $\mathcal{M}$ is defined. We cannot reproduce the exact relations because of the lack of steady-state isotropic driven turbulence and "Fourier diffusivity" analysis. We can still perform qualitative comparisons. In a $5 \kpc$ box for our simulation, we find that}

\begin{ceqn}
\begin{align}\label{eq:DC_vdis}
    \langle \mathrm{DC} \rangle \propto \kappa \propto \spread^2/t \propto \vdis^n.
\end{align}
\end{ceqn}

\rev{From \autoref{sec:diffusion_coefficient}, we know that $n\approx1$; thus, $\kappa \propto \mathcal{M}$, same as \citet{Colbrook2017}. And our time-dependent part $\spread^2/t$ is potentially captured by their scale-dependent part $(k\mathrm{L_{box}})^{-\alpha_{c}}$ from Fourier diffusivity analysis. Despite our inability to determine $\alpha_c$ as $\vdis$ and $\spread$ are not independent, our results emphasise the need for a dynamic diffusivity model (scale- and/or time-dependent) even in the case of transonic CGM turbulence ($\mathcal{M}$ ranges from $\sim 0.2-1.2$ in our simulation).}

\rev{Finally, for the numerical aspects of mixing,} the standard + $2 \kpc$ run has the lowest resolution and the same amount of dye injection as the highest resolution $1 \kpc$ run with eight cells injection. After the initial jump due to numerical reasons in \autoref{fig:num_mix} (bottom panel), both standard + $1 \kpc$ runs have the same rate of $\spread$ increase, whereas the standard + 2 kpc run shows an increase nearly double that. This again suggests that gas mixing at the resolution of our simulations is significantly affected by numerical diffusion. \rev{However,} since we perform our study at a fixed spatial resolution, the dependence of gas mixing on the gas properties is likely not affected, even if the amount of mixing is. We explicitly tested the correlation results with standard + $2 \kpc$ resolution simulations and found qualitatively the same results.

\begin{figure}
    \includegraphics[width=1\columnwidth]{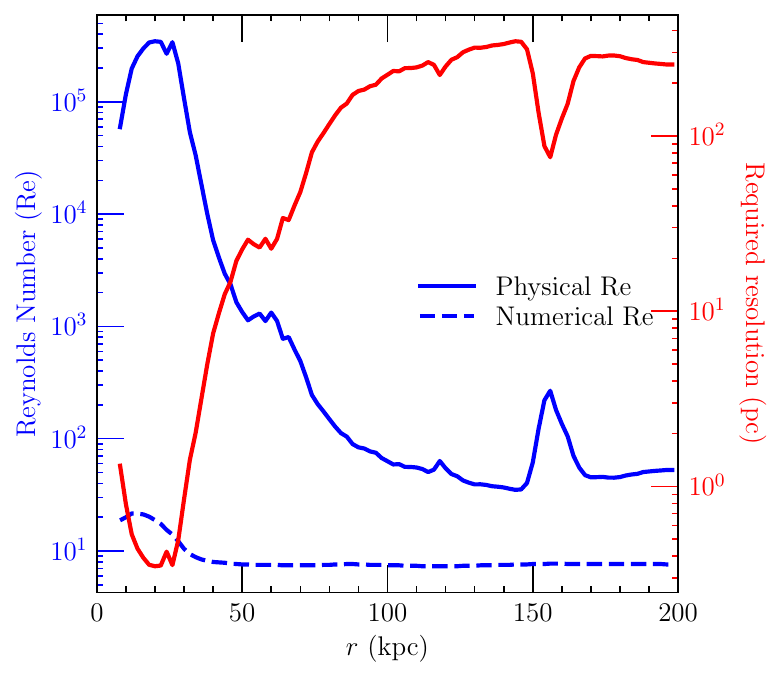}
    \caption{Left axis (blue) shows the physical $\rm Re$ (solid) and numerical $\rm Re$ (dashed), and the right axis (red) shows the resolution required for the physical $\rm Re$ to be equal to numerical $\rm Re$ (essentially, resolve turbulence) as a function of distance from the centre ($r$). The small peak/dip at $\sim 150 \kpc$ is due to the presence of a satellite galaxy. This graph suggests that we need a few hundred pc to resolve outer CGM turbulence beyond $\sim 100 \kpc$ (current resolution $\sim$ 1 \kpc).}
    \label{fig:Re_analysis}
\end{figure}

\subsection{Reynolds numbers for CGM turbulence}
The numerical resolution of a simulation is critical for accurately capturing turbulence, which also affects gas mixing. To better understand the limitations of current simulations and their ability to capture turbulence, we calculate the physical and numerical Reynolds numbers ($\rm Re$) at different distances from the centre ($r$) of the galaxy. The kinetic Reynolds number is defined as ${\rm Re}=\vdis \, l_{\rm D}/\nu$, where $\vdis$ is the velocity dispersion on a characteristic length scale of $l_{\rm D}=5\kpc$ (see \autoref{fig:spread_time_order} and \autoref{fig:vdis_ld}), and $\nu$ is the kinematic viscosity \citep[calculated from Eq.~3.16 in][assuming fully collisional gas]{Brandenburg_2005} \rev{with the following equation:}

\begin{ceqn}
\begin{align}
    \nu = 1.3 \times 10^{23}\left(\frac{T}{10^6~\mathrm{K}}\right)^{5/2}\left(\frac{n_i}{\mathrm{cm}^{-3}}\right)^{-1} \mathrm{cm}^2\mathrm{s}^{-1}.
\end{align}
\end{ceqn}

The volume-weighted \rev{average} densities and temperatures \rev{in a $5 \kpc$ box} are used to calculate $\nu$ at a given $r$. Typical $\rm Re$ values for the inner ($50<r/\kpc<75$) and outer ($75<r/\kpc<200$) CGM are $\sim10^3$ and $10^2$, respectively (solid blue curve in \autoref{fig:Re_analysis})\rev{, when we use the velocity dispersion \rev{at $5 \kpc$ scales} (see \autoref{fig:vdis_ld})}. For Kolmogorov turbulence, the \rev{effective} numerical ${\rm Re} \approx N_{\rm grid}^{4/3}$ considering numerical viscosity \citep{Schmidt2015, Kriel2022, Shivakumar2025}, where $N_{\rm grid}$ is the number of cells along one dimension in the turbulent region under consideration. \rev{We assume this because in subsonic turbulence, $\vdis \propto l^{1/3}$; thus, ${\rm Re}\propto l^{4/3}/\nu$. All the energy at grid scale is dissipated (${\rm Re}=1$), which gives us $\nu \propto \Delta x^{4/3}$, where $\Delta x$ is the grid resolution. The effective numerical ${\rm Re}\propto N_{\rm grid}^{4/3}$ follows from this, where the proportionality constant can be assumed to be $\approx 1$ \citep{Shivakumar2025}.} Thus, if a cell has a size of $\sim 1 \kpc$, which is roughly the case here, we have only 5 cells in the region, and the numerical $\rm Re$ in that region is expected to be $\sim 5^{4/3} \sim 8$ (dashed blue curve in \autoref{fig:Re_analysis}). \rev{The effective numerical ${\rm Re}$ quantifies the ability a simulation to resolve the physical ${\rm Re}$. When the effective numerical $\rm Re$ of a simulation is equal to the physical $\rm Re$, the numerical dissipation scale is the same as the viscous dissipation scale, and the turbulent cascade is modelled accurately.}

\autoref{fig:Re_analysis} shows that physical $\rm Re$ is always greater than numerical $\rm Re$, indicating that our simulations are numerically viscous. This leads to a loss of turbulent structures, as the dissipation occurs on larger scales than expected. The red curve shows the required resolution to match physical $\rm Re$ with the numerical $\rm Re$. Resolution better than $1 \kpc$ is needed to capture the full turbulent cascade in the simulations. While it may not be possible to capture the inner CGM turbulence in MW-mass galaxies (the resolution required is a few tens of pc or higher) in cosmological simulations, it is possible to capture turbulence in the outer CGM ($\ge 75 \kpc$) with a resolution of a few hundred pc. This resolution may be where the convergence for gas mixing lies, because when we improve the resolution further, the effects of numerical viscosity would be less than the physical turbulent mixing. Understanding the limitations of current simulations and improving their accuracy in capturing turbulence could help to correctly interpret the gas properties, specifically the mixing and distribution of metals, which also affects gas cooling and gas accretion onto the central galaxy. Even though numerical diffusion is significant in our simulations, we have checked that our correlation results from \autoref{sec:mag_gas_mixing} and \autoref{sec:dir_gas_mixing} remain valid at all resolutions. 

\section{Conclusions}
\label{sec:conclusions}
In this study, we explore the mechanisms responsible for gas mixing in the CGM using cosmological, magnetohydrodynamical simulations \rev{conducted with the moving-mesh \textsc{arepo} code}. Most of our analysis is performed with simulations that have standard mass refinement plus a minimum spatial resolution of $1 \kpc$. We insert tracer dyes (passive scalars) in 95 locations representing diverse CGM environments -- interacting inflow-outflow pairs, pure (coherent) inflows, pure (coherent) outflows, static gas, and outflows interacting with static flows. We track the evolution of the dye for $200 \Myr$ to understand the nature of mixing. Our findings are summarised below.

\begin{enumerate}
    \item Interacting inflow-outflow pairs show a higher level of gas mixing with the surroundings than the pure coherent inflows and outflows. The static gas experiences similar levels of gas mixing to the pure coherent inflows and outflows. 

    \item To infer which gas property causes gas mixing, we perform the Spearman rank correlation test between the dye spread and local gas properties such as temperature, metallicity, density, kinetic energy, magnetic field, velocity dispersion, \rev{symmetric shear tensor (or strain-rate tensor) magnitude, and antisymmetric shear tensor (or vorticity tensor) magnitude}. We find that the velocity dispersion \rev{and traceless symmetric shear tensor (or pure shear deformation) magnitude \citep[][see \autoref{sec:mag_gas_mixing} for more details]{Smagorisnky1963}} correlate the best with dye spread. Thus, random and/or shearing motions likely cause gas mixing. \rev{Although not all SPH simulations include a subgrid model for metal diffusion, some SPH simulations estimate diffusivity using either velocity dispersion or symmetric shear tensors.} Weaker, but still significant correlations are also found with other gas properties like density and kinetic energy, likely due to an underlying correlation with velocity dispersion.

    \item When we correlate the dye spread after $200 \Myr$ with \rev{local gas properties} at different times in different-sized boxes around the dye location, we find that the correlation is maximal for velocity dispersion and \rev{symmetric shear tensor magnitudes} measured in smaller regions ($\leq 5 \kpc$) around dye injection locations $\sim 70-120 \Myr$ after dye injection. This implies that the impact of \rev{these velocity-derived quantities} on gas mixing is largest after a $\sim 100 \Myr$ delay. We find that the diffusion coefficient calculated from dye spread has a nearly linear dependence on the velocity dispersion with a power-law exponent of $1.1 \pm 0.09$ (robust to changes in resolution). \rev{Scaling constants range from $0.1-0.6$ when estimating diffusion using velocity dispersion, and the Smagorinsky constants vary from $0.2-0.4$ when estimating diffusion from shear tensors.} These numerical simulation results from implicit modelling of gas mixing can be useful to estimate diffusion in explicit subgrid models employed by some SPH simulations \citep{Klessen2003, Greif2009, Williamson2016}.

    \item \rev{Our alignment analysis reveals that velocity and magnetic field dispersion both correlate with the dyed cloud’s shape. However, vectors calculated from the traceless symmetric shear tensor and antisymmetric shear tensor exhibit a stronger alignment. Specifically, the dispersion of stretching (deformation) eigenvectors of the symmetric shear tensors and plane-of-rotation eigenvectors of the antisymmetric shear tensors shows the maximum alignment with the dye shape.} Nearly \rev{70} per cent of our \rev{dye} sample has \rev{stretching eigenvectors' dispersion} and dye spread \rev{direction} aligned to within \rev{9.6} degrees. The shape of the dye spread correlates best with \rev{gas properties} measured in larger boxes ($\ge 10 \kpc$) \rev{contrary to the correlation of the extent of mixed dye, which is larger in smaller boxes ($0-5 \kpc$). Overall, we find that traceless symmetric shear tensors best predict the extent and shape of gas mixing in our simulations, slightly better than velocity dispersion.}

    \item We find that numerical mixing plays an important role in our simulations. When we change the refinement criterion of fixed spatial resolution from $1 \kpc$ to $2 \kpc$, the dye spreads twice as quickly. However, despite the amount of gas mixing changing, the strong qualitative correlation with velocity dispersion remains. 

    \item The velocity dispersion decreases with increasing galactocentric radius. This implies a reduction in gas mixing further out in the CGM, useful for understanding metal diffusion \citep{Nugent2025}. Velocity dispersion can be used to compute Reynolds numbers in the CGM. The inner ($50<r/\kpc<75$) and outer ($75<r/\kpc<200$) CGM have typical values of $\sim10^3$ and $10^2$, respectively. We find that the numerical resolution is not enough to resolve the physical Reynolds numbers. In order to resolve turbulence in the outer CGM, the required resolution would be a few hundred $\pc$ for the outer CGM ($r>75$~kpc) and a few tens of $\pc$ or smaller for the inner CGM.

    \item The dye spread exhibits a power law dependence with time, having most exponents ranging from $\sim 0.7-1.1$. In normal (Brownian) diffusion, a square root time dependence (exponent $\sim 0.5$) is expected. Thus, we have a case of superdiffusion or hyperballistic diffusion over $200 \Myr$ timescales, which could be caused by turbulence, large-scale coherent flows, and numerical diffusion in the CGM. Thus, the diffusion coefficient is a time-varying quantity in our simulations.

\end{enumerate}

Despite the inability of our cosmological simulations to capture the full cascade of turbulence in the CGM, a fixed spatial resolution in the CGM aids us in filtering out the numerical effects on gas mixing. Although the amount of gas mixing is not converged with resolution, the differences we found in gas mixing in different physical environments were found to be robust. Thus, our findings that the amount and the direction of gas mixing correlate the best with velocity dispersion \rev{and symmetric shear tensors} hold, irrespective of the resolution. These results help us understand the fundamental nature of gas and metal diffusion in the CGM, which is crucial to understanding the gas flows into and out of galaxies and the distribution of metals in the CGM. The topic of multiphase turbulent mixing is complex and many unanswered questions remain, such as \rev{the impact of magnetic fields}, and whether we can fully resolve CGM turbulence and obtain converged results. We plan to tackle these questions in the future.

\section*{Acknowledgements}
\rev{We would like to thank our referee, Douglas Rennehan, for detailed, useful comments that greatly improved the quality of the manuscript.} We would like to thank Deovrat Prasad, Andrew Hannington, Andrew Cook, Thomas Rintoul, Mark Krumholz, Neco Kriel, Alex Ji, Ayan Acharyya, and James Neeson for the valuable discussion that greatly improved the quality of this study.

FvdV is supported by a Royal Society University Research Fellowship (URF\textbackslash R1\textbackslash 191703 and URF\textbackslash R\textbackslash241005).

\rev{AS acknowledges support from the Australian Research Council's Discovery Early Career Researcher Award (DECRA, project DE250100003).}

CF acknowledges funding provided by the Australian Research Council (Discovery Project DP230102280), and the Australia-Germany Joint Research Cooperation Scheme (UA-DAAD). CF further acknowledges high-performance computing resources provided by the Leibniz Rechenzentrum and the Gauss Centre for Supercomputing (grants~pr32lo, pr48pi and GCS Large-scale project~10391), the Australian National Computational Infrastructure (grant~ek9) and the Pawsey Supercomputing Centre (project~pawsey0810) in the framework of the National Computational Merit Allocation Scheme and the ANU Merit Allocation Scheme.

This work used the DiRAC Memory Intensive service (Cosma8, under project code dp221) at Durham University, managed by the Institute for Computational Cosmology on behalf of the STFC DiRAC HPC Facility (www.dirac.ac.uk). The DiRAC service at Durham was funded by BEIS, UKRI and STFC capital funding, Durham University and STFC operations grants. DiRAC is part of the UKRI Digital Research Infrastructure.

The authors gratefully acknowledge the Gauss Centre for Supercomputing e.V. (\url{https://www.gauss-centre.eu}) for funding this project
(with project code pn68ju) by providing computing time on the GCS Supercomputer SuperMUC-NG at Leibniz Supercomputing Centre (\url{https://www.lrz.de}).

The research made use of Auriga \citep{GrandEA2017}, NumPy \citep{Harris2020}, matplotlib \citep{Hunter2007}, SciPy \citep{Virtanen2020}\rev{, and Meshoid \citep{grudic_meshoid_2022}} software packages.

\section*{Data Availability}

Data related to this work will be shared on reasonable request to the corresponding author.



\bibliographystyle{mnras}
\bibliography{refs.bib} 




\appendix

\section{Interacting Inflow-Outflow Selection Methodology}
\label{sec:appendix_io}
 To find interacting inflows-outflows ($io$), first, we look for all inflow and outflow \rev{voxels} (see \autoref{fig:binned_h12_class}) whose centres are within $20 \sqrt{3} \kpc$ ($2 \times$ \rev{voxel} diagonal) of one another, where \rev{voxels} represent $10 \times 10 \times 10 \kpc$ gas regions. This is done to narrow down the adjacent (potentially interacting) \rev{voxels} as we aim to find two regions of gas that are likely to interact in the near future. We then locate the points on both \rev{voxel} centre trajectories where the closest approach occurs. As the shortest distance is perpendicular to both trajectories, it can be found using
\begin{ceqn}
\begin{align}
    \vec{\prone}+t_1 \vec{\pVone} + t_3 (\vec{\pVone} \times \vec{\pVtwo}) = \vec{\prtwo} + t_2 \vec{\pVtwo}, 
\end{align}
\end{ceqn}
where 1 and 2 denote an inflow and outflow \rev{voxel}, respectively, $\vec{r_{1/2}^p}$ are their position vectors, $\vec{V_{1/2}^p}$ are their velocity vectors. The equation can be solved for $t_1, t_2$, and $t_3$. As the interaction will happen in the direction of the respective velocity vectors, only positive values of $t_1$ and $t_2$ are considered valid. The points of the closest approach also have to be near the initial \rev{voxel} positions on the simulation timescale of $200 \Myr$. Thus, we make sure that the distance between the initial \rev{voxel} positions and the respective points of closest approach is smaller than the distance a \rev{voxel} would travel in $150 \Myr$ at the current speed. Furthermore, the distance between the points of closest approach has to be smaller than $10 \kpc$, i.e. the \rev{voxel} size.

\begin{figure}
\includegraphics[width=1\columnwidth]{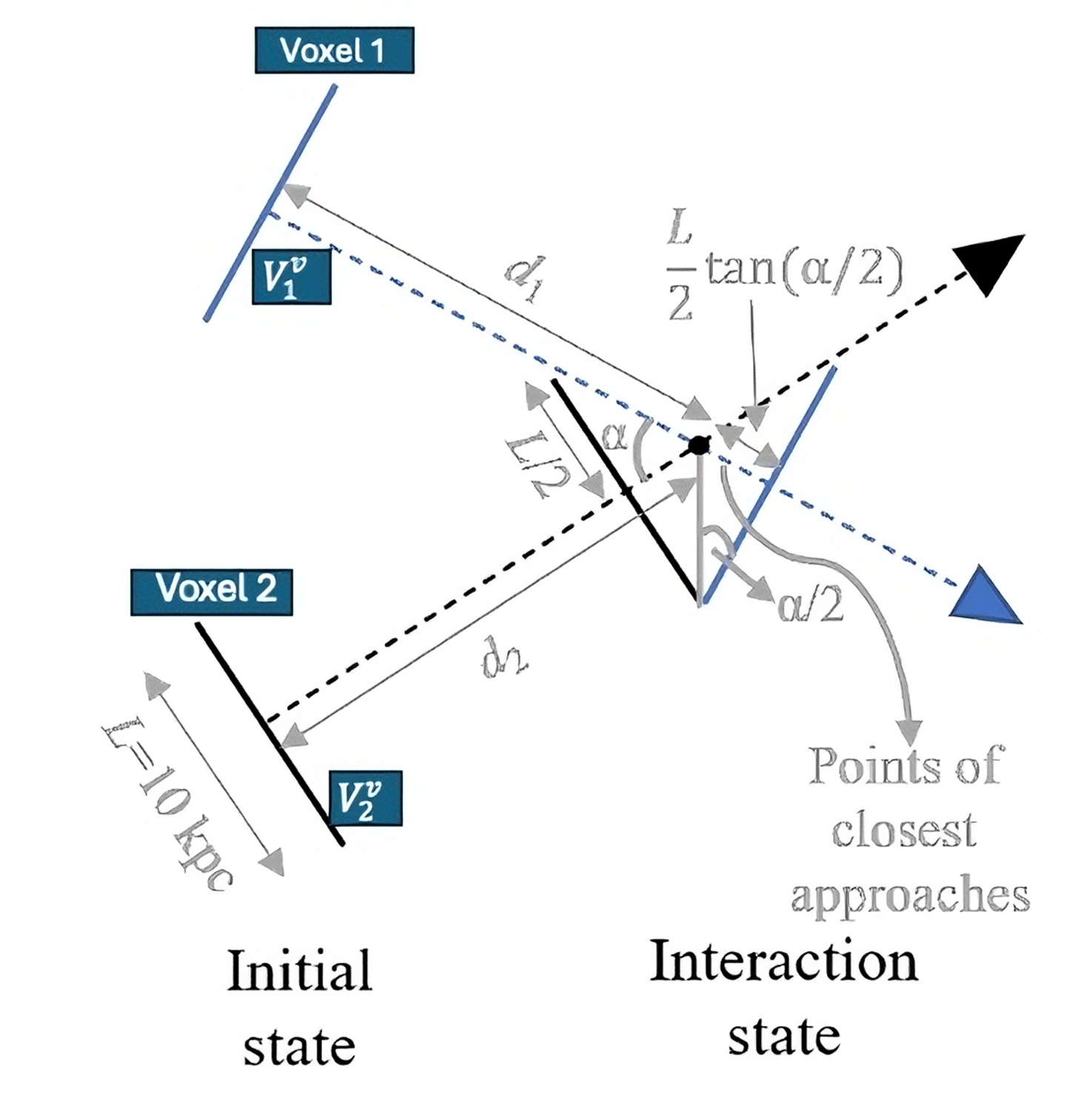} 

\caption{Top-view (along $\pVone \times \pVtwo$) schematic of two \rev{voxels} (shown as 1D \rev{voxel} cross-sections) represented by a blue and a black line ($10 \times 10 \times 10 \kpc$) in the CGM headed towards each other. The interaction state shown here is an extreme case of \rev{voxel} interaction (where \rev{voxels} barely overlap) after assuming their velocities to remain constant. Blue and black colours are used with \rev{voxels} 1 and 2, respectively. Points of the closest approach, overlapping from the top view, represent the shortest distance between the trajectories of both \rev{voxels}.}
\label{fig:poca_geo}
\end{figure}

Lastly, to ensure interaction, both \rev{voxels} must reach their respective points of closest approach nearly simultaneously. \autoref{fig:poca_geo} shows one extreme case where an interaction between the \rev{voxels} is considered to be just possible (bottom edges of the faster blue \rev{voxel} and slower black \rev{voxel} just touch). In this extreme case, the time taken from the initial to interaction state (traced out from trajectories of \rev{voxels} assuming constant velocities) would be the same for both \rev{voxels}, which can be expressed as

\begin{ceqn}
\begin{align}\label{eq:poca_time1}
    \frac{d_1+ (L / 2)~{\rm \tan(\alpha/2)}}{\pVone} = \frac{d_2 - (L / 2)~{\rm \tan(\alpha/2)}}{\pVtwo},
\end{align}
\end{ceqn}
where $d_1$ and $d_2$ are distances from initial \rev{voxel} positions to \rev{voxel} positions at the point of closest approach, $L=10 \kpc$ is the width of a \rev{voxel}, and $\alpha$ is the angle between the \rev{voxel} trajectories. \autoref{eq:poca_time1} can be expressed as 
\begin{ceqn}
\begin{align}\label{eq:poca_time2}
    \left| \frac{d_1}{\pVone}-\frac{d_2}{\pVtwo} \right| = \frac{L}{2} {\rm \tan(\alpha/2)}\left(\frac{1}{\pVone}+\frac{1}{\pVtwo}\right)
\end{align}
\end{ceqn}
The left-hand side (LHS) of this equation represents the absolute difference in time taken by the \rev{voxels} to reach their respective points of closest approach. Thus, the two \rev{voxels} could interact if the LHS is smaller than the right-hand side.

\section{Gas Property and Resolution Test Plots}
\label{sec:appendix_misc}
\begin{figure}
    \includegraphics[width=1\columnwidth]{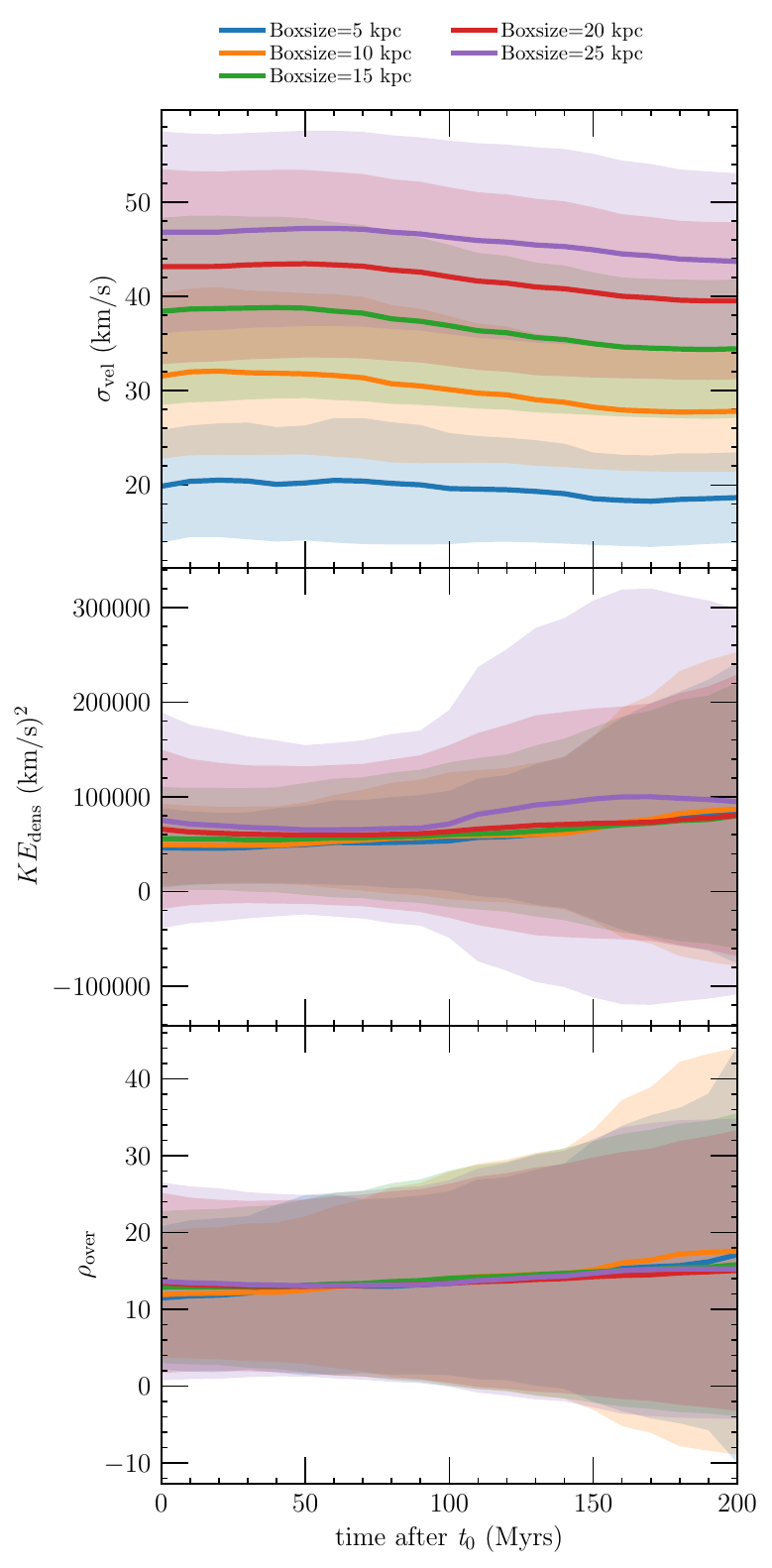} 
    \caption{Gas properties averaged over different boxes (with varying sizes) around the dye injection locations as a function of time after injection. The solid lines and the shaded regions around them show the mean and the standard deviation of the dataset, respectively. The velocity dispersion remains constant with time, but scales to the power of $\sim 0.5$ with box size. Mean kinetic energy density and overdensity remain roughly constant with time and box size. Despite all gas properties varying minimally with time, the gas property exhibiting direct correlation with dye spread might still have varying correlation strengths temporally.}
    \label{fig:init_prop_time}
\end{figure}

\autoref{fig:init_prop_time} shows the mean (solid curves) and standard deviation (shaded regions) of the velocity dispersion, kinetic energy density, and overdensity around all the dyes in regions of different sizes at different times. Velocity dispersion scales with box size to the power $0.5$ while kinetic energy density and overdensity remain constant. The mean remains constant with time while the scatter increases for kinetic energy density and overdensity. This shows that some dyes undergo major changes with time, but the general environmental character remains constant when averaging over all cases.

\begin{figure}
    \includegraphics[width=1\columnwidth]{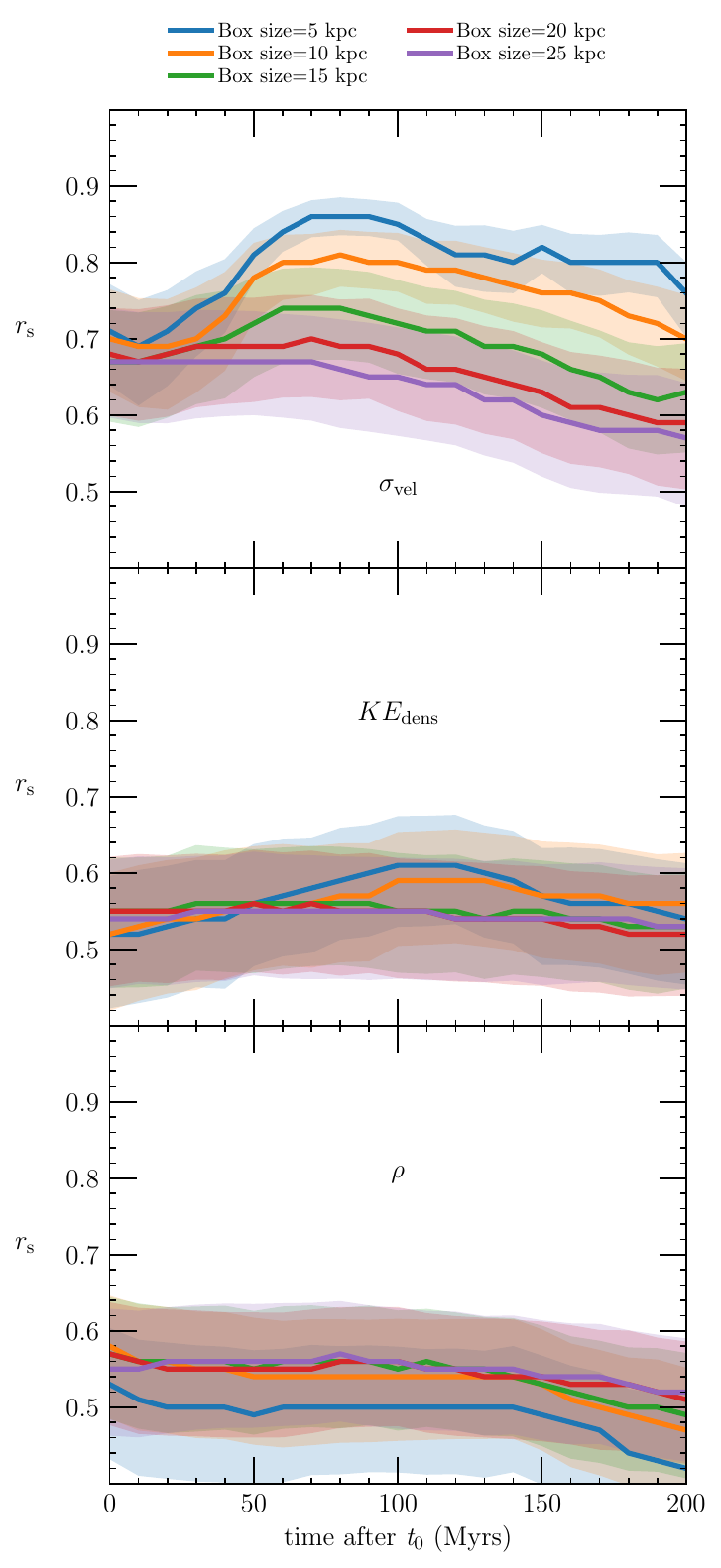} 
    \caption{Spearman rank coefficient ($\rs$) of the correlation between the magnitude of dye spread after $200 \Myr$ and averaged gas properties in boxes around the dye injection locations with different sizes at different times. The errorbars represent $1\sigma$ confidence intervals from bootstrapping the sample 10,000 times. This is the same graph as \autoref{fig:spread_time_order} but with an eight-cell dye injection at each location instead of one. }
    \label{fig:clump_corr}
\end{figure}

\autoref{fig:clump_corr} shows the Spearman rank ($\rs$) correlation coefficient between dye spread and velocity dispersion, kinetic energy density, and overdensity averaged over different box sizes around the dyes at different times. The dye is initially injected in clumps of eight cells for \autoref{fig:clump_corr} while it was injected in single cells for \autoref{fig:spread_time_order}. The velocity dispersion still correlates best with the spread of the dye (maximum $\rs \sim 0.86$) and the results remain very similar to one-cell injection. The peak in $\rs$ is attained slightly earlier, however, possibly due to dye spreading out to optimal configuration for transfer of velocity dispersion to spread faster.


\bsp	
\label{lastpage}
\end{document}